\pdfoutput=1
\documentclass[sigconf, nonacm, pdfa]{acmart}
\usepackage[utf8]{inputenc}

\newcommand\vldbdoi{10.14778/3749646.3749725}

\newcommand\vldbavailabilityurl{https://github.com/BillyZhaohengLi/SIEVE-vldb25}
\newcommand\vldbpagestyle{empty}

\usepackage{subfiles} 
\usepackage{bm}
\usepackage[linesnumbered,ruled,noend]{algorithm2e}
\usepackage{algorithmic}
\usepackage{hyperref}
\usepackage{cleveref}
\usepackage{tcolorbox}
\crefformat{section}{§#2#1#3}
\Crefformat{section}{§#2#1#3}
\crefname{figure}{Fig}{Figs}
\Crefname{figure}{Fig}{Figs}
\crefname{table}{Table}{Tables}
\Crefname{table}{Table}{Tables}
\Crefname{algorithm}{Alg.}{Algs.}
\usepackage{xspace}
\usepackage{subfiles} 
\usepackage{multirow}
\usepackage{tabularx}
\usepackage{subcaption}
\usepackage{pgfplots}
\usepackage{pgfplotstable}
\pgfplotsset{compat=1.18} 
\usepackage{ifthen}
\usepackage[shortlabels]{enumitem}
\setlist[enumerate]{leftmargin=0.5cm,topsep=0.5mm}
\setlist[itemize]{leftmargin=0.5cm,topsep=0.5mm}
\usepackage{tikz}
\usepackage{wasysym}
\usetikzlibrary{calc}
\usetikzlibrary{fit}
\usetikzlibrary{patterns}
\usetikzlibrary{decorations.pathreplacing}
\usetikzlibrary{arrows,calc,arrows.meta}
\usetikzlibrary{matrix,positioning}
\usetikzlibrary{shapes.geometric}

\usepackage{makecell}
\usepackage{pifont}

\newcommand{\diskann}{{\sf DiskANN}\xspace}
\newcommand{\spann}{{\sf SPANN}\xspace}

\newcommand{\gist}{{\sf GIST1M}\xspace}
\newcommand{\deep}{{\sf DEEP10M}\xspace}
\newcommand{\msspace}{{\sf MSSPACE10M}\xspace}
\newcommand{\bigann}{{\sf BIGANN1B}\xspace}

\definecolor{PreFilterColor}{HTML}{4D4D4D}
\definecolor{HnswBaseColor}{HTML}{dadada}
\definecolor{SmartHnswBaseColor}{HTML}{F95454}
\definecolor{OracleColor}{HTML}{6D6D6D}
\definecolor{ParlayIvfColor}{HTML}{6439FF}
\definecolor{CapsColor}{HTML}{38A528}
\definecolor{DiskAnnColor}{HTML}{6439FF}
\definecolor{AcornGammaColor}{HTML}{03045E}
\definecolor{AcornOneColor}{HTML}{0D92F4}
\definecolor{OursColor}{HTML}{C62E2E}
\definecolor{GreenColor}{HTML}{38A528}

\definecolor{ReadColor}{HTML}{cf3457}
\definecolor{WriteColor}{HTML}{ffb570}

\definecolor{vintagegreen}{HTML}{ABDDA4}

\definecolor{GreedyColor}{HTML}{7081ff}
\definecolor{RandomColor}{HTML}{ffb570}
\definecolor{NoneColor}{HTML}{6D6D6D}

\definecolor{Redborder}{HTML}{805861}
\definecolor{Greenborder}{HTML}{384180}
\definecolor{Blueborder}{HTML}{566F52}
\definecolor{Greyborder}{HTML}{4D4D4D}
\definecolor{Lightgrey}{HTML}{dadada}
\definecolor{ExampleColor1}{HTML}{7081ff}
\definecolor{ExampleColor2}{HTML}{ffb0c2}
\definecolor{Lightred}{HTML}{ffb09c}
\definecolor{Lightblue}{HTML}{b8e2f2}
\definecolor{FlagColor}{HTML}{CCCCCC}

\definecolor{vintageblue}{HTML}{7081ff}
\definecolor{vintagered}{HTML}{721817}

\definecolor{NoOptColor}{HTML}{264653}
\definecolor{LRUColor}{HTML}{777777}
\definecolor{RandomColor}{HTML}{2a9d8f}
\definecolor{GreedyColor}{HTML}{e9c46a}
\definecolor{HeuristicColor}{HTML}{f4a261}
\definecolor{SCColor}{HTML}{e76f51}
\definecolor{AllColor}{HTML}{CCCCCC}

\definecolor{SAColor}{HTML}{ffb0c2}
\definecolor{SeparatorColor}{HTML}{9b5de5}
\definecolor{BlueColor}{HTML}{0081a7}

\newcommand{\midsepremove}{\aboverulesep = 0.3mm \belowrulesep = 0.3mm}
    \midsepremove
    \newcommand{\midsepdefault}{\aboverulesep = 0.605mm \belowrulesep = 0.984mm}
    \midsepdefault

\begin{document}

%%
%% The "title" command has an optional parameter,
%% allowing the author to define a "short title" to be used in page headers.
\title[Cloud-Native Vector Search: A Comprehensive Performance Analysis]{Cloud-Native Vector Search: A Comprehensive Performance Analysis [Experiments, Analysis and Benchmark]}

%%
%% The "author" command and its associated commands are used to define
%% the authors and their affiliations.
%% Of note is the shared affiliation of the first two authors, and the
%% "authornote" and "authornotemark" commands
%% used to denote shared contribution to the research.
% \author{Temp Author}
% \email{temp_author@bytedance.com}
% \affiliation{%
%   \country{USA}
%   \institution{ByteDance Inc.}
% }
\author[Zhaoheng Li, Wei Ding, Silu Huang, Zikang Wang, Yuanjin Lin, Ke Wu, Yongjoo Park, Jianjun Chen]{Zhaoheng Li$^*$, Wei Ding$^{\dagger}$, Silu Huang$^{\dagger}$, Zikang Wang$^{\dagger}$, Yuanjin Lin$^{\dagger}$, Ke Wu$^{\dagger}$, Yongjoo Park$^*$, Jianjun Chen$^{\dagger}$}
% \authornote{Work done during internship at Bytedance.}
\affiliation{%
  \country{ 
     University of Illinois Urbana-Champaign$^*$ \quad Bytedance$^{\dagger}$}
}
\email{{zl20,yongjoo}@illinois.edu, {wei.ding,silu.huang, wangzikang, linyuanjin, ke.wu, jianjun.chen}@bytedance.com}

% \author{Zhaoheng Li}
% \authornote{Work done during internship at Bytedance.}
% \affiliation{%
%   \country{UIUC}
% }
% \email{zl20@illinois.edu}

% \author{Silu Huang}
% \affiliation{%
%   \country{Bytedance Inc.}
% }
% \email{silu.huang@bytedance.com}

% \author{Wei Ding}
% \affiliation{%
%   \country{Bytedance Inc.}
% }
% \email{wei.ding@bytedance.com}

% \author{Yongjoo Park}
% \affiliation{%
%   \country{UIUC}
% }
% \email{yongjoo@illinois.edu}

% \author{Jianjun Chen}
% \affiliation{%
%   \country{Bytedance Inc.}
% }
% \email{jianjun.chen@bytedance.com}
%%
%% By default, the full list of authors will be used in the page
%% headers. Often, this list is too long, and will overlap
%% other information printed in the page headers. This command allows
%% the author to define a more concise list
%% of authors' names for this purpose.
% \renewcommand{\shortauthors}{Trovato et al.}

%%
%% The abstract is a short summary of the work to be presented in the
%% article.
\begin{abstract}
Vector search has been widely employed in recommender system and retrieval-augmented-generation pipelines, commonly performed with \emph{vector indexes} to efficiently find similar items in large datasets. Recent growths in both data and task complexity have motivated placing vector indexes onto remote storage---\textit{cloud-native vector search}, which cloud providers have recently introduced services for.
Yet, despite varying workload characteristics and various available vector index forms, providers default to using \emph{cluster-based} indexes, which on paper do adapt well to differences between disk and cloud-based environment: their fetch granularities and lack of notable intra-query dependencies aligns with the large optimal fetch sizes and minimizes costly round-trips (i.e., as opposed to \emph{graph-based} indexes) to remote storage, respectively.

This paper systematically studies cloud-native vector search: \emph{What and how should indexes be built and used for on-cloud vector search}? We analyze bottlenecks of two common index classes, cluster and graph indexes, on remote storage, and show that despite current standardized adoption of cluster indexes on the cloud, graph indexes are favored in workloads requiring high concurrency and recall, or operating on high-dimensional data or large datatypes. We further find that on-cloud search demands significantly different indexing and search parameterizations versus on-disk search for optimal performance. Finally, we incorporate existing cloud-based caching setups into vector search and find that certain index optimizations \textit{work against} caching, and study how this can be mitigated to maximize gains under various available cache sizes.

\end{abstract}
% \silu{mention what we sacrifice and what we have leveraged, e.g., 2x memory and with 25\% workload. }.
% \silu{maybe use medium/average instead of upto}

\maketitle
%% The following content must be adapted for the final version
% paper-specific

% %%% do not modify the following VLDB block %%
% %%% VLDB block start %%%
\pagestyle{\vldbpagestyle}
% \begingroup\small\noindent\raggedright\textbf{PVLDB Reference Format:}\\
% Zhaoheng Li, Silu Huang, Wei Ding, Yongjoo Park, Jianjun Chen. \vldbtitle. PVLDB, \vldbvolume(\vldbissue): \vldbpages, \vldbyear.\\
% \href{https://doi.org/\vldbdoi}{doi:\vldbdoi}
% \endgroup
% \begingroup
% \renewcommand\thefootnote{}\footnote{\noindent
% This work is licensed under the Creative Commons BY-NC-ND 4.0 International License. Visit \url{https://creativecommons.org/licenses/by-nc-nd/4.0/} to view a copy of this license. For any use beyond those covered by this license, obtain permission by emailing \href{mailto:info@vldb.org}{info@vldb.org}. Copyright is held by the owner/author(s). Publication rights licensed to the VLDB Endowment. \\
% \raggedright Proceedings of the VLDB Endowment, Vol. \vldbvolume, No. \vldbissue\ %
% ISSN 2150-8097. \\
% \href{https://doi.org/\vldbdoi}{doi:\vldbdoi} \\
% }\addtocounter{footnote}{-1}\endgroup
% %%% VLDB block end %%%

% %%% do not modify the following VLDB block %%
% %%% VLDB block start %%%F
% \ifdefempty{\vldbavailabilityurl}{}{
% \vspace{.3cm}
% \begingroup\small\noindent\raggedright\textbf{PVLDB Artifact Availability:}\\
% The source code, data, and/or other artifacts have been made available at \url{\vldbavailabilityurl}.
% \endgroup
% }
% %%% VLDB block end %%%

%%
%% The code below is generated by the tool at http://dl.acm.org/ccs.cfm.
%% Please copy and paste the code instead of the example below.
%%

%%
%% This command processes the author and affiliation and title
%% information and builds the first part of the formatted document.

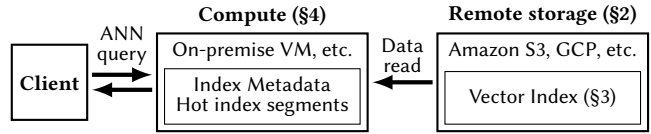
\begin{figure}[t]

\tikzset{
codenode/.style={
    draw=black,minimum width=16mm,
    font=\small\ttfamily,
},
cellnode/.style={
    font=\small\sffamily,
    anchor=south west,inner ysep=0,
},
technode/.style={
    draw=black,text width=34mm,
    align=left,
    font=\small\sffamily,
}
}

\centering
\begin{tikzpicture}

\def\g{0.6}

% user interface
\node[draw=black,thick,anchor=north west,
    minimum width=10mm,minimum height=10mm,align=center,
    font=\small\bfseries] 
    (client) at (0,0) {Client};

\node[draw=black,thick,anchor=west,
    minimum width=28mm,minimum height=13mm,align=center,
    font=\small\sffamily] 
    (A1) at ($(client.east)+(0.9, 0)$) {};
\node[anchor=south,font=\small\bfseries] 
    at ($(A1.north)+(0,-0.02)$)
    {Compute (\cref{sec:caching})};
\node[anchor=north,font=\small\sffamily] 
    at ($(A1.north)+(0,-0.02)$)
    {On-premise VM, etc.};
\node[anchor=south,draw=black,    font=\small\sffamily,minimum width=26mm, minimum height=5mm,align=center] 
    (C1) at ($(A1.south)+(0, 0.1)$)
    {Index Metadata\\[-0.3em] Hot index segments};
% \node[anchor=south,font=\small\bfseries] 
%     at ($(C1.north)+(0,-0.02)$)
%     {Local Cache (\cref{sec:caching})};

% kernel
\node[draw=black,thick,anchor=north west,
    minimum width=28mm,minimum height=13mm,align=center,
    font=\small\sffamily] 
    (B3) at ($(A1.north east)+(0.9,0)$) {};
\node[anchor=south,font=\small\bfseries] 
    at ($(B3.north)+(0,-0.02)$)
    {Remote storage (\cref{sec:background})};
\node[anchor=north,font=\small\sffamily] 
    at ($(B3.north)+(0,-0.02)$)
    {Amazon S3, GCP, etc.};
\node[anchor=south,draw=black,    font=\small\sffamily,minimum width=26mm, minimum height=7mm,align=center] 
    (I1) at ($(B3.south)+(0, 0.1)$)
    {Vector Index (\cref{sec:cloud_index})};
% \node[anchor=south,draw=black,    font=\small\sffamily,minimum width=26mm, minimum height=5mm,align=center] 
%     (I2) at ($(I1.north)+(0, 0.1)$)
%     {Graph indexes (\cref{sec:graph_index})};

\node[anchor=south,align=center,font=\small\sffamily] 
    at ($(A1.east)+(0.45, 0.02)$)
    {Data\\[-0.3em]read};
\node[anchor=south,align=center,font=\small\sffamily] 
    at ($(client.east)+(0.45, 0.12)$)
    {ANN\\[-0.3em]query};
    
% Arrows
\draw[-latex, ultra thick] 
    ($(client.east)+(0.05,0.1)$) -- ($(A1.west)+(-0.05,0.1)$);
\draw[latex-, ultra thick] 
    ($(client.east)+(0.05,-0.1)$) -- ($(A1.west)+(-0.05,-0.1)$);
% \draw[-latex, ultra thick] 
%     ($(B2.east)+(0.05,0.2)$) -- ($(B3.west)+(-0.05,0.2)$);
\draw[latex-, ultra thick] 
    ($(A1.east)+(0.05,0)$) -- ($(B3.west)+(-0.05,0)$);

\end{tikzpicture}

\vspace{-3mm}
\caption{Cloud-native vector search setup we study in this paper. Compute fetches vector index segments from remote storage and utilizes local resources to cache hot segments.}
\label{fig:intro}
\vspace{-3mm}
\end{figure}
\section{Introduction}
\label{sec:intro}
    % Obligatory introduction to vector search \begin{enumerate}
    %     \item Modern ML models produce high-quality embeddings
    %     \item Used for recommender systems and RAG
    %     \item Due to large data volumes, vector indexes are used for approximate search
    %     \item As datasets grow even larger, indexes no longer fit in memory and on-disk indexes (e.g., DiskANN, SPANN) have been studied
    % \end{enumerate}

Finding semantically similar items via \textit{vector search} is a cornerstone of many recent recommender systems and retrieval-augmented-generation (RAG) pipelines~\cite{lian2020lightrec, mathur2024vector, wang2024searching, rusum2024vector}.
The problem of doing so efficiently, with \textit{vector indexes}, has gained popularity as usage of large-scale vector datasets~\cite{bigann1b, biganngithub} become feasible via efficient embedding models~\cite{llama, phi, openai}.
Oftentimes, depending on task specifications, the index may reside in different locations: it may need to be \textit{in-memory} for applications demanding sub-millisecond latency (e.g., fraud detection~\cite{frauddetection}), while tasks focused on cost-effectiveness (e.g., eCommerce~\cite{li2018design}) may prefer indexes to be \textit{on-disk}.

\paragraph{Emergence of Cloud-Native Vector Search}
Recently, as data scales demanded by vector search grow even larger (e.g., Bing Visual Search~\cite{visualsearch} and EBay Recommendations'~\cite{ebayvector} billion-scale datasets), providers such as TurboPuffer~\cite{turbopuffer} and Amazon S3 Vector~\cite{amazons3vector} have began offering cloud-native vector search services, providing cost-efficient solutions for storing and querying these massive vector datasets by storing them and their indexes on remote storage. 
\cref{fig:intro} depicts a typical pipeline: vector indexes are stored as (immutable) \textit{objects} on remote storage, then, clients interacts with \textit{compute} which performs ANN queries by fetching index \textit{segments} for computation, while using local resources (e.g., memory, SSD) for caching (e.g., storing metadata of hot indexes~\cite{turbopufferarchitecture}).

% Apart from traditional recommendation-based tasks (e.g., Bing Visual Search~\cite{visualsearch} and EBay Recommendations~\cite{ebayvector}) which feature large, billion-scale vector datasets too large to be stored on disk, recent RAG and Agentic AI~\cite{turbopufferagentic} tasks feature a significantly different form of large data scale---many small-to-medium vector datasets (up to million-scale~\cite{amazons3vector}), each owned and queried by a subset of users and/or applications (e.g., timestamped data slices), with a large combined size. \cref{fig:intro} depicts a typical pipeline: vector indexes are stored as (immutable) \textit{objects} on remote storage, then, the client interacts with \textit{compute} which performs ANN queries by fetching index \textit{segments} for computation while utilizing optimizations such as caching (e.g., storing metadata of hot indexes~\cite{turbopufferarchitecture}).

\paragraph{Current Providers Default to Cluster Indexes}
There currently exists two distinct commonly-used index classes for vector search: \textit{cluster indexes} based on inverted lists for text search~\cite{zobel2006inverted} which organize datasets into KMeans-clustered posting lists, and \textit{graph indexes}, which build KNN-like graphs over datasets and serve queries via iterative graph traversal. Despite the wide variety of workloads with different characteristics and requirements (e.g., dataset size, recall, latency, and throughput), almost all current on-cloud vector search providers offer only cluster indexes (e.g., IVF-PQ~\cite{ivfpq}, SPANN~\cite{chen2021spann}, SPFresh~\cite{xu2023spfresh}). In principle, cluster indexes' search characteristics align with recent findings on remote storage performance~\cite{gadban2021analyzing, hou2017understanding, durner2023exploiting}: Data is read at posting list (often tens of KBs in size) granularity from cluster indexes which matches the large (8-16MiB) cost/throughput-optimal fetch sizes, and the typical lack of intra-query dependencies---all required posting lists can be fetched at once---minimizes long roundtrip times (millisecond level) to remote storage. Yet, prior studies for on-disk indexing, which is highly similar with on-cloud search interface-wise (i.e., compute fetches required index \textit{blocks} from disk), have shown that some workloads significantly favor graph indexes, with notable examples being high recall and/or throughput scenarios~\cite{cheng2024characterizing, renstorage}.

\paragraph{Our Goal: How to Optimize Cloud-Native Vector Search?}
Hence, we hypothesize that unlike the one-size-fits-all cluster index-based approach of current providers, different workloads and setups call for different indexing choices, which we study in this paper: \emph{What and how should vector indexes be built and used for cloud-native vector search?} We analyze bottlenecks faced by cluster and graph indexes for on-cloud search to answer the following sub-questions: 

\textbf{(RQ1) What index for what scenario?} We aim to identify the optimal index class given dataset characteristics and query throughput, recall, latency, concurrency requirements. We show that graph indexes could be favored in high concurrency and/or recall scenarios (e.g., Agentic AI tasks~\cite{turbopufferagentic}) and vice versa (e.g., ad-hoc recommendations~\cite{visualsearch}). More interestingly, low dataset dimensions (e.g., 96D GIST descriptors~\cite{gist}) and datatypes (e.g., INT8~\cite{msspace}) shift the balance in favor of graph indexes. 

\textbf{(RQ2) How to design indexes?} Cluster and graph indexes can be built in ways that vary in characteristics such as the amount of computation and data read per query. We aim to identify the optimal parameterization based the remote storage's characteristics (e.g., I/O bandwidth). We find that, for cloud-native vector search, parameterization should prioritize reducing I/O, even if this substantially increases compute costs (such as by using more fine-grained posting lists for graph indexes). Notably, such configurations are often documented as suboptimal for on-disk indexing~\cite{diskannpy, chen2021spann}.

\textbf{(RQ3) How to utilize caching?} Caching for vector search is relatively understudied, with existing providers~\cite{turbopuffer} and works~\cite{zhou2025govector} defaulting to maximizing cache hit rate as proxy for performance. We study \textit{how} indexes benefit from caching, and interestingly find that some indexing optimizations diminish caching gains (e.g., \diskann's multi-neighbor expansion), where higher cache hit rate does not directly lead to lower latency: Optimal indexing parameterizations (e.g., \spann's replication factor~\cite{chen2021spann}) can heavily depend on cache size, and certain sizes may call for these optimizations to be \textit{tuned down} to maximize caching gains for on-cloud search.

\paragraph{Contributions}
According to our observations on the characteristics of existing and hypothetical cloud-native vector search pipelines (\cref{sec:background}), we contribute the following with our benchmark study:
\begin{enumerate}
    \item \textbf{Index Design (\cref{sec:cloud_index}):} We follow our theory-driven modeling and bottleneck analysis of graph and cluster vector indexes for cloud-native vector search to formulate hypotheses on optimal indexing designs and parameterizations.
    \item \textbf{Index-Cache Integration (\cref{sec:caching}):} We explore integrating caching methods of prior on-disk indexing and adjacent cloud analytics works into cloud-native vector search, identify missed opportunities, and hypothesize on how to improve caching gains.
    \item \textbf{Experimental Evaluation (\cref{sec:experiments}):} We perform extensive evaluation on 4 real-world vector datasets on realistic on-cloud and on-disk setups. We provide actionable insight on \textit{what index to use} and \textit{how to use them} in a wide variety of scenarios differing in recall/throughput/latency requirements, query concurrency, dataset characteristics, network resources, and cache size.
\end{enumerate}
\section{Cloud Vector Search Fundamentals}
\label{sec:background}

Recent \textit{cloud-native} vector search architectures feature significant differences versus prior vector search works utilizing cloud storage~\cite{guo2022manu, opensearch}. We will describe the key characteristics of this emerging framework in \cref{sec:background_vector}, then, its architectural bottlenecks in \cref{sec:background_characteristics}, then, the bottlecks' implications for on-cloud vector search in \cref{sec:cloud_index_cost}.

\subsection{Cloud-Native Architecture}
\label{sec:background_vector}
Cloud-based vector search setups have been explored by a variety of providers such as Manu~\cite{guo2022manu}, Milvus~\cite{wang2021milvus}, and OpenSearch~\cite{opensearch}. These works follow the commonly used practice in prior shared-architecture DBMSes~\cite{yang2018intermediate, michiardi2019memory} and visualization dashboards~\cite{zgraggen2016progressive, eichmann2020idebench} of completely separating compute and storage: indexes are built and held on dedicated compute nodes for querying~\cite{guo2022manu}, and are streamed to remote storage for persistence. While this approach offers high ad-hoc query performance as indexes are queried in-memory, costs can be high when large-scale datasets are involved.

\paragraph{Object Storage-First: Significant Cost Savings}
Accordingly, TurboPuffer~\cite{turbopuffer} and Amazon S3 Vector~\cite{amazons3vector} have proposed the object storage-first, cloud-native vector search architecture that focuses on cost savings for large-scale workloads. These frameworks treat remote storage (e.g., Amazon S3~\cite{amazons3}, Azure Blob~\cite{azureblob}, GCP~\cite{gcp}, and Volcano TOS~\cite{volcenginetos}) as the truth source on which indexes are stored. Then, instead of querying indexes in-memory, they follow similar procedures as querying on-disk indexes (e.g., \diskann~\cite{diskannpy}), where required index \textit{segments} (i.e., analogus to disk-based \textit{blocks}, e.g., posting lists for the SPFresh index~\cite{xu2023spfresh} utilized by TurboPuffer~\cite{turbopuffer}) are fetched for querying.
Then, local resources on compute nodes (e.g., memory, SSD) can be used as a cache for hot index segments, for example, TurboPuffer currently caches the cluster metadata (namely the BKT tree, \cref{sec:cluster_index_cost}) of commonly accessed SPFresh indexes.
While this approach in general results in higher query latency due to incurring roundtrips to remote storage for fetching data that may be cold, operation costs are reduced (10-100$\times$\cite{turbopuffer}) as indexes no longer need to be in-memory: storing indexes in remote storage is significantly cheaper ($\sim$20\$ / TB / month~\cite{amazons3pricing, volcenginetos}).

\subsection{Cloud Environments: Constrained I/O}
\label{sec:background_characteristics}
Despite remarkably similar procedures between querying indexes on disk and remote storage---existing disk-based indexes such as \diskann and \spann can be queried \textit{as is} when uploaded to remote storage (e.g., via mounting APIs~\cite{azuremounting}), the environment differences significantly affect performance of cloud-native vector search (\cref{fig:exp_cpu_usage}).
This section describes I/O characteristics of remote storage that constrain cloud-native vector search performance. We present characteristics of the remote storage (Volcano TOS~\cite{volcenginetos}) and local SSD we use for experiments (\cref{sec:experiments}) in this paper for reference in \cref{tbl:env_diff}.

% Object stores such as Amazon S3~\cite{amazons3}, Azure Blob~\cite{azureblob}, GCP~\cite{gcp}, and Volcano TOS~\cite{volcenginetos} act a cost-efficient way to separate storage from compute, offering significantly cheaper storage costs ($\sim$20\$ / TB / month~\cite{amazons3pricing, volcenginetos}) versus VM-attached SSDs ($\sim$100\$ 
%  / TB / month~\cite{ssdpricing}).

 % while on-disk querying may be CPU-bound (e.g., high concurrency and recall \diskann workload, \cref{fig:exp_cpu_usage_diskann}), on-remote storage querying is often network-bound: two common disk-based indexes, \diskann (\cref{fig:diskann_overhead_offcpu}) and \spann (\cref{fig:spann_overhead_offcpu}) are both blocked by data reads while CPU remains underutilized.

\paragraph{Low Read Throughput}
Read throughput from remote storage to the compute node is determined primarily by two factors: ingress bandwidth of compute controlled by its network interface card, and the throughput limit of the remote storage prefix holding the vector index.\footnote{All existing cloud-native vector search providers use one-compute-node-to-one-storage-bucket setups. We defer studying distributed setups to future work.} Providers often limit the latter by throttling I/O due to shared infrastructure~\cite{amazons3}, with the specific limit often depending on compute location---on-premise machines use \textit{internal network}, which typically has higher throughput (e.g., up to 50Gbps~\cite{volcenginebandwidth, amazons3bandwidth}), while machines that access through \textit{external network} are subject to lower throughputs (e.g., up to 5Gbps~\cite{volcenginebandwidth, hou2017understanding}). In general, remote storage throughput is significantly lower versus local SSDs (up to 15GB/s~\cite{haas2023modern}); hence, disk-based indexes may exhibit throughput bottlenecks on remote storage (\cref{sec:cluster_index_cost}).

\begin{table}[t]
\caption{Key differences of the specific remote storage and SSD we use for experimentation (\cref{sec:experiments}) in this paper.}
\vspace{-3mm}
\footnotesize
\addtolength{\tabcolsep}{-2pt} 
\begin{tabular}{l rrr}
\toprule
\textbf{Storage} & \textbf{p50 Read Latency} & \textbf{GET QPS}  & \textbf{Read Throughput (GB/s)} \\
 \midrule
SSD & 66.5$\mu$s & 420,000 & 12 \\
Volcano TOS~\cite{volcenginetos} & 9000$\mu$s & 20,000 & 0.625 \\

\bottomrule
\end{tabular}
\vspace{-3mm}
\addtolength{\tabcolsep}{2pt}
\label{tbl:env_diff}
\end{table}

\paragraph{Long Read Latency}
The read latency of fetching data from remote storage to the compute node is determined by the hotness of the data segment being fetched: cold data (e.g., ad-hoc queries) incurs latency of 30-200ms, while hot, frequently-accessed data (e.g., long-tailed accesses of TurboPuffer's Agentic AI workloads~\cite{turbopufferagentic}) incurs latency as low as 10ms~\cite{su2019understanding, durner2023exploiting}.
Even for hot data, these numbers are still significantly larger than the 10-100$\mu$s of recent SSDs~\cite{lee2019asynchronous, wang2024sindex}, which degrades the performance of disk-based indexes that incur multiple roundtrips during search on remote storage (\cref{sec:graph_index_cost}).

\paragraph{Low Rate Limits}
Cloud providers often limit GET request frequency issued to each prefix (i.e., each index), ranging from 5,500 (S3~\cite{amazons3limit}) to 20,000 (Azure~\cite{azurebloblimit} and TOS~\cite{volcenginebandwidth}) queries-per-second (QPS)~\cite{shue2012performance}. This limit is lower than typical modern SSDs (>100K~\cite{moshayedi2008enterprise}), and acts as a second limit to on-cloud search throughput alongside bandwidth, especially in high-concurrency settings with smaller data scales (empirically observed in \cref{sec:exp_e2e}). While this can potentially be addressed by caching hot index segments with local resources, current providers only cache index \textit{metadata}~\cite{turbopuffer}, leading to missed opportunities; we explore integrating generalized and vector index-specific caching strategies into cloud-native vector search in \cref{sec:caching}.

\begin{figure}[t]\captionsetup[subfigure]{font=footnotesize}
\pgfplotsset{scaled y ticks=false}
\centering
\begin{subfigure}[b]{\linewidth}
\centering
\caption{Most significant overhead categories of \spann indexing configurations on-disk on the \gist dataset with nprobe=8 and concurrency=1.}
\footnotesize
\midsepremove
\begin{tabular}{|l|c|c|c|}
\toprule
 % & \multicolumn{16}{c|}{Indexing Method}\\
 %  \midrule
Index Location & I/O\% & Distance Comps \% & BKT Tree\% \\
 \hline
Disk & 31.23 & \textbf{51.23} & \textbf{7.88} \\
\hline
Remote Storage & \textbf{54.16} & 28.99 & 6.10 \\
% \hline
% \# Exact searches@$efc=10$ & \textbf{3907} & 3821 & \textbf{2697} & 2099  \\
\bottomrule
\end{tabular}
\midsepdefault
\label{tbl:spann_overhead_metrics}
\end{subfigure}
\hfill
\begin{subfigure}[b]{\linewidth}
\centering
\caption{Most significant overhead categories of \diskann indexing configurations on-disk on the \gist dataset with search\_len=10 and concurrency=1.}
\footnotesize
\midsepremove
\begin{tabular}{|l|c|c|c|}
\toprule
 % & \multicolumn{16}{c|}{Indexing Method}\\
 %  \midrule
Index Location & I/O\% & Memory Operations \% & PQ Distance Comps.\% \\
 \hline
Disk & 68.69 & 11.20 & 7.80 \\
\hline
Remote Storage & \textbf{70.74} & 11.27 & 11.13 \\
% \hline
% \# Exact searches@$efc=10$ & \textbf{3907} & 3821 & \textbf{2697} & 2099  \\
\bottomrule
\end{tabular}
\midsepdefault
\label{tbl:diskann_overhead_metrics}
\end{subfigure}
\vspace{-1mm}
\caption{Key overheads of \spann and \diskann on \gist measured with \texttt{perf}~\cite{linuxperf}. Both indexes' search costs are dominated by I/O, albeit different aspects of it, on remote storage.
}
\vspace{-3mm}
\label{tbl:cloud_index_overhead}
\end{figure}

\subsection{On-Cloud Search Costs: Dominated by I/O}
\label{sec:cloud_index_cost}
This section describes search cost models of two common vector index classes---cluster and graph indexes, and how constrained I/O of cloud environments may affect their on-cloud performance.

\subsubsection{Cluster Index Search Cost}
\label{sec:cluster_index_cost}
Cluster indexes are based on inverted lists for text search~\cite{zobel2006inverted, brown1994fast}: the vector dataset is K-means clustered into $n$ \textit{posting lists} represented by \textit{centroids}~\cite{jegou2011searching}.
For search, a number of posting lists (controlled by parameter \texttt{nprobe}) with centroids closest to the query vector are fetched (e.g., \spann uses a balanced K-means tree (BKT) built on centroids), and data vectors in the posting lists are compared with the query vector to find top-$k$ results.
Cost can be modeled for an index $C$ as follows:
% involves three steps: finding posting lists, fetching data, and performing distance computations, which can be modeled for a cluster index $C$ as follows:
\begin{equation}
    C_{cluster}(C) = c_{centroid}(n,nprobe)+c_{fetch}(l)+l*c_{dist}
\label{eq:cluster}
\end{equation}
Where $c_{centroid}(n,nprobe)$ is the cost of finding the top \texttt{nprobe}-out-of-$n$ posting lists, $c_{fetch}(l)$ is the cost of fetching $l$ total vectors in the posting lists from storage, and $l*c_{dist}$ is the cost for $l$ distance computations with the query vector. Cluster indexes features no dependencies between posting list fetches, which is a major advantage given the long roundtrips to remote storage (\cref{sec:background_characteristics}): resource-allowing, all \texttt{nprobe} posting lists can be fetched at once.

\paragraph{Cluster Indexes' Throughput Bottleneck} Higher query recalls demand higher \texttt{nprobe} values, which linearly increases $c_{fetch}(l)$ and $l*c_{dist}$, while $c_{centroid}(n,nprobe)$ often becomes increasingly negligible (e.g., \spann's BKT trees have $O(nlog(nprobe))$ scaling~\cite{tavallali2021k}).
Hence, cluster indexes's performance is limited by which of I/O (for $c_{fetch}(l)$) or compute's (for $l*c_{dist}$) bottleneck is reached first: We show via \spann, a commonly-used cluster index, has on-cloud search costs that are dominated by the former (\cref{tbl:spann_overhead_metrics}), which is also the bottleneck for its cloud-native vector search performance as we observe in \cref{tbl:spann_overhead_metrics}. This bottleneck is especially prominent under high query concurrencies and recalls (\cref{fig:teaser_io_usage_spann}): for example, \spann, reading large amounts of data per query at high recalls~\cite{chen2021spann} (empirically verified in \cref{fig:exp_index_metrics_dataread} where \spann reads \textit{256MB} data per 0.995 recall \gist query), when combined with high query concurrencies, is bottlenecked by \textit{data not arriving frequently enough} (\cref{fig:teaser_io_usage_spann}) while compute is underutilized (\cref{fig:teaser_cpu_usage_spann}). Hence, reducing data read is a key optimization point for cluster indexes' on-cloud performance, which we study in \cref{sec:cloud_index_improve}.

% : for example, \spann, which reads large amounts of data per query at high recalls~\cite{chen2021spann} (empirically verified in \cref{fig:exp_index_metrics_dataread} where \spann reads \textit{256MB} data per 0.995 recall \gist query), when combined with high query concurrencies, is bottlenecked by \textit{data not arriving frequently enough} (\cref{fig:teaser_io_usage_spann}) while compute is underutilized (\cref{fig:teaser_cpu_usage_spann}). We formally analyze this bottleneck and explore potential solutions in \cref{sec:cluster_index_cost}.

\input{sections/plots/background_cpu_usage}
\newlength{\textfloatsepsave} 
\setlength{\textfloatsepsave}{\textfloatsep} \setlength{\textfloatsep}{-4mm}
\newlength{\floatsepsave} 
\setlength{\floatsepsave}{\floatsep} \setlength{\floatsep}{-4mm}
\begin{algorithm}[t]
% \small
\caption{\small{Graph-based iterative search for ANN} \label{alg:graph}}
\SetKw{KwIn}{Input:}
\SetKw{KwOut}{Output:}
\footnotesize
\KwIn{query vector $q$, entry points $ep$, result size $k$, cand. set size $search\_len$}\\
\KwOut{$k$ closest vectors to $q$}\\
Initialize visited set $V=\emptyset$, candidate set $C=ep$; \\
\While{$C\backslash V \neq \emptyset$}{
  $c \leftarrow $ extract nearest vector in $C$ to $q$\\
  $C \leftarrow C \cup neighborhood(c)$\\
  $R \rightarrow R\cup \{c\}$\\
  \textbf{if} $|C|>search\_len$:\\
  $| \,\,\,$update $C$ to keep $search\_len$ closest points to $q$\\
}
\textbf{Return} $k$ closest points to $q$ in $C$.
\end{algorithm}
\setlength{\textfloatsep}{\textfloatsepsave}
\setlength{\floatsep}{\floatsepsave}
\subsubsection{Graph Index Search Cost}
\label{sec:graph_index_cost}
Graph indexes are based on KNN graphs, with each data vector connected to at most $K$ close neighbors~\cite{malkov2018efficient, fu2017fast, jayaram2019diskann}.\footnote{Different notations are used for indegree limit $K$, e.g., DiskANN's $R$ and HNSW's $M$.}
Search is performed via iterative graph traversal from \textit{entry points} while maintaining a candidate set of current closest neighbors (with size bounded by \texttt{search\_len}); then, top-$k$ results are retrieved from the candidate set post-traversal (\cref{alg:graph}). 
While current graph indexes often feature some unique optimizations (e.g., HNSW's multi-layer structure~\cite{malkov2018efficient} and \diskann's extracting multiple vectors at once from the candidate set~\cite{jayaram2019diskann}), graph search cost can be largely modeled for an index $G$ as follows:

\begin{equation}
    C_{graph}(G) = rt\times(TTFB + c_{fetch}(K) + K\times c_{dist})
\label{eq:graph_cost}
\end{equation}

Where $rt$ is the number of traversal iterations (line 4 in \cref{alg:graph}), $TTFB$ (i.e., time-to-first-byte) is read latency to storage, $c_{fetch}(K)$ is cost of fetching each round's neighborhood from storage, and $K\times c_{dist}$ is the per-round vector distance computation cost. 
Notably, despite parallelizable data reads and distance computations in each round, the $rt$ expansion rounds must be performed iteratively; that is, even if bandwidth and compute were \textit{unlimited}, search still at minimum incurs $rt\times TTFB$ cost for communicating with storage.

\paragraph{Graph Indexes' Read Latency Bottleneck}
Higher query recalls demand the usage of larger candidate set sizes, which incurs more roundtrips, especially noticeable for high-recall queries (\cref{sec:background_characteristics}).
Even for on-disk search, prior studies have observed that graph indexes' search latencies are dominated by these incurred roundtrip times~\cite{wang2024starling}. Accordingly, for \diskann, a commonly-used graph index, we observe in \cref{tbl:diskann_overhead_metrics} that I/O\% (i.e., $TTFB$) plays an even larger role in query latency due to remote storages' longer read latencies (\cref{sec:background_characteristics}): for example, ad-hoc querying with \diskann,\footnote{High-concurrency \diskann workloads can also be bottlenecked by IOPS (\cref{sec:exp_parameterization}).} which requires long, iterative traversals for high recalls~\cite{jayaram2019diskann} (empirically verified in \cref{fig:exp_index_metrics_roundtrips} where \diskann performs \textit{43} roundtrips per 0.995 recall \gist query), is bottlenecked by \textit{waiting for data} (\cref{fig:diskann_overhead_offcpu}) while \textit{both} compute (\cref{fig:teaser_cpu_usage_diskann}) and I/O bandwidth (\cref{fig:teaser_io_usage_diskann}) are underutilized.  Hence, reducing roundtrips is a key optimization point for graph indexes' on-cloud performance, which we study in \cref{sec:cloud_index_improve}.

% for example, ad-hoc querying with \diskann,\footnote{High-concurrency \diskann workloads can also be bottlenecked by IOPS (\cref{sec:cloud_index_improve}).} which requires long, iterative traversals for high recalls~\cite{jayaram2019diskann} (empirically verified in \cref{fig:exp_index_metrics_roundtrips} where \diskann performs \textit{43} roundtrips per 0.995 recall \gist query), is bottlenecked by \textit{waiting for data} (\cref{fig:diskann_overhead_offcpu}) while \textit{both} compute (\cref{fig:teaser_cpu_usage_diskann}) and I/O bandwidth (\cref{fig:teaser_io_usage_diskann}) are underutilized. We formally analyze this bottleneck and explore solutions in \cref{sec:graph_index_cost}.

\paragraph{RQ1: What Index for What Scenario?}
Given the bottlenecks that we have formulated and observed for two commonly-used cluster and graph indexes, \spann and \diskann, we can hypothesize that one reason cloud providers opt for cluster indexes is because their respective bottleneck---read throughout, degrades less from local disk to remote storage versus graph indexes' read latency (10-20$\times$ versus $\sim$150$\times$), leading to lower performance drops (\cref{fig:teaser_qps_spann}). Yet, our observations also hint at cases where graph indexes may outperform cluster indexes: for example, in \cref{fig:teaser_qps_diskann}, individual queries' underutilization of I/O bandwidth leads to its performance scaling remarkably well w.r.t. query concurrency; we perform a detailed study of \textit{when should graph indexes be used over cluster indexes for cloud-native search, and vice versa?} in \cref{sec:exp_e2e}.

\section{Index Design for Cloud Storage}
\label{sec:cloud_index}
\label{sec:cloud_index_improve}
% As we have shown in \cref{sec:background}, \spann and \diskann, two representative on-disk cluster and graph indexes, are bound by different aspects of I/O on remote storage. We aim to describe these bottlenecks in the context of their search cost models in \cref{sec:cloud_index_cost}, then, hypothesize on how they can be tuned for better on-cloud performance in \cref{sec:cloud_index_improve}.
As we have shown both empirically and via cost modeling in \cref{sec:background}, two common vector index classes, cluster and graph indexes, are bound by different aspects of I/O on remote storage. This section presents some potential solutions to tuning these indexes better on-cloud performance.
\paragraph{Reducing Cluster Indexes' Data Read Per Query}
One of \spann's main optimizations for reducing data read per query is to duplicate boundary vectors (i.e., vectors with similar distance to multiple centroids) up to a user-specified \texttt{num\_replica} times. While this increases individual posting list (and overall index) sizes (by up to 3$\times$ versus IVF~\cite{zobel2006inverted}, \cref{sec:exp_parameterization}), it reduces the \texttt{nprobe} value to reach target recalls, and is often a net reduction in total data read and distance computations~\cite{chen2021spann}.
This optimization, along with \spann's increased size being a minor factor due to cheap remote storage costs (\cref{sec:background_vector}), has lead it to be TurboPuffer's index of choice~\cite{turbopufferarchitecture}.\footnote{TurboPuffer uses SPFresh~\cite{xu2023spfresh}, an in-place updatable extension of \spann.}

Orthogonal to replication, we can increase centroid count ($n$ in \cref{eq:cluster}) to build an index with more posting lists each containing fewer vectors. Doing so would trade more distance computations for less I/O (\cref{tbl:spann_index_tuning}); while this is discouraged for on-disk use due to both increasing BKT tree search costs~\cite{chen2021spann} and index sizes (\cref{tbl:spann_index_size}), such an index can improve on-cloud search performance where the former is dominated by $c_{fetch}(l)$ (verified in \cref{sec:exp_parameterization}).
\begin{figure}[t]\captionsetup[subfigure]{font=footnotesize}
\pgfplotsset{scaled y ticks=false}
\centering
\begin{subfigure}[b]{\linewidth}
\centering
\caption{Most significant overhead categories of \spann indexing configurations on the \gist dataset with nprobe=640 and concurrency=1.}
\footnotesize
\midsepremove
\begin{tabular}{|l|c|c|c|}
\toprule
 % & \multicolumn{16}{c|}{Indexing Method}\\
 %  \midrule
Index & I/O\% & Distance Comps.\% & BKT Tree \% \\
 \hline
\spann (default centroid\%=16) & \textbf{63.76} & 4.03 & 1.61 \\
\hline
\spann (centroid\%=32) & 55.30 & \textbf{4.46} & \textbf{4.46} \\

% \hline
% \# Exact searches@$efc=10$ & \textbf{3907} & 3821 & \textbf{2697} & 2099  \\
\bottomrule
\end{tabular}
\midsepdefault
\label{tbl:spann_index_tuning}
\end{subfigure}
\hfill
\vspace{-1mm}
\begin{subfigure}[b]{\linewidth}
\centering
\caption{Most significant overhead categories of \diskann indexing configurations on the \gist dataset with search\_len=640 and concurrency=1.}
\footnotesize
\midsepremove
\begin{tabular}{|l|c|c|c|}
\toprule
 % & \multicolumn{16}{c|}{Indexing Method}\\
 %  \midrule
Index & I/O\% & Memory Ops. \% & PQ Distance Comps.\% \\
 \hline
\diskann (default R=64) & \textbf{80.00} & 10.00 & 10.00 \\
\hline
\diskann (R=256) & 71.42 & \textbf{14.29} & \textbf{14.29} \\
% \hline
% \# Exact searches@$efc=10$ & \textbf{3907} & 3821 & \textbf{2697} & 2099  \\
\bottomrule
\end{tabular}
\midsepdefault
\label{tbl:diskann_index_tuning}
\end{subfigure}
\vspace{-3mm}
\caption{Overheads of on-cloud \spann and \diskann. Indexes can be tuned to reduce I/O for more computations.
}
\vspace{-4mm}
\label{tbl:index_tuning}
\end{figure}

\paragraph{Reducing Graph Indexes' Roundtrip Count} \diskann utilizes the technique of extracting multiple vectors (controlled by \textit{beamwidth} $W$~\cite{diskannpy}) at once from the candidate set per expansion round (\cref{sec:graph_index_cost}),\footnote{Each iteration is not guaranteed to branch into the full $W$ points, e.g., when there are fewer than $W$ unvisited nodes in the candidate set~\cite{diskannzilliz, jayaram2019diskann}.} effectively performing branching traversal for each roundtrip as opposed to point-at-a-time of prior memory-based indexes' (line 5 in \cref{alg:graph}, e.g., HNSW~\cite{malkov2018efficient} and NSG~\cite{fu2017fast}). This trades more total I/O calls per query for lower roundtrip count, and is often set to a moderate value (e.g., 4-16~\cite{diskannpy}) to balance read latency and IOPS limits (\cref{sec:background_characteristics}) of the environment and workload characteristics, and should be carefully tuned: using a high $W$ value can degrade search by reaching the IOPS limit first (observed in \cref{fig:exp_diskann_larger_beamwidth_concurrency_qps}). 

Alternatively, to avoid increasing \textit{any} I/O-related metrics, a denser graph with higher $K$ in (\cref{eq:graph_cost}) can be used, which instead trades more distance computations and larger graph size for fewer I/O calls per query \textit{and} roundtrips. Like fine-grained cluster indexes, while dense graphs are discouraged on disk, they can often lead to performance gains for on-cloud search (\cref{tbl:diskann_index_tuning}) due to distance computations being relatively significant (verified in \cref{sec:exp_parameterization}).

\paragraph{(RQ2) How to build indexes?} Through our cost modeling in terms of cluster and graph (specifically, \spann and \diskann) indexes' parameters, we have identified how notable parameters can be adjusted to control the trade-offs between costs (e.g., I/O, compute). We perform benchmarking to find \textit{How should cluster and graph indexes be built and used for optimal on-cloud performance?} in \cref{sec:exp_parameterization}.

\section{On-Cloud Index-Cache Integration}
\label{sec:caching}

Existing cloud-native search providers cache index metadata with local memory and SSD resources on compute: specifically, TurboPuffer caches the BKT tree of commonly-accessed \spann indexes (\cref{sec:background}). Doing so does benefit queries on those indexes, and has also been explored in prior on-disk works such as \diskann and IVF-PQ~\cite{ivfpq}'s in-memory caching of product-quantized dataset and codebook, respectively. Yet, following our bottleneck analyses in \cref{sec:cloud_index_cost}, the currently unexplored approach of additionally caching index data may bring further gains: this section explores existing caching works for cloud native vector search (\cref{sec:caching_background}), and discuss integration of caching and vector index optimizations in (\cref{sec:caching_benefit}).

\begin{figure}[t]\captionsetup[subfigure]{font=footnotesize}
\pgfplotsset{scaled y ticks=false}
\centering

\begin{subfigure}[b]{0.49\linewidth}
\centering
\begin{tikzpicture}

\begin{axis}[
    xtick=data,
    width=38mm,
height=28mm,
    ymin=0,
    ymax=1.5,
    clip=false,
    axis y line*=none,
    axis x line*=none,
    tick align=outside,
    ytick={0, 0.5,1.0,1.5},
    yticklabels={0, 0.5,1.0,1.5},
    xlabel=Data read (MB),
    xlabel style={yshift = 1.5ex},
    label style={font=\scriptsize},
        ylabel style={yshift=-1ex,xshift=-0.5ex, font=\scriptsize},
    xmin = 0,
    xmax = 260,
    xtick={0, 50, 100, 150, 200, 250},
    xticklabels={0, 50, 100, 150, 200, 250},
    x tick label style={yshift=0.5ex},
    tick label style={font=\scriptsize},
    legend style={
        at={(-0.2,1.1)},anchor=south west,column sep=2pt,
        draw=black,fill=white,
        inner ysep=0.1pt,
        /tikz/every even column/.append style={column sep=5pt},
        font=\footnotesize
    },
    legend cell align={left},
    legend columns=5,
    ylabel={Query latency (s)},
    ymajorgrids,
    every axis plot/.append style={thick}
    % legend image code/.code={%
    % \draw[#1, draw=none] (0cm,-0.1cm) rectangle (0.6cm,0.1cm);}
]

% \addplot[line width=1pt, GreenColor,mark=*] coordinates
% {(1, 51.2) (2, 47.7) (3, 50.5) (4, 62.2) (5, 63.0)(6, 75.2)};
% Read
\node[anchor=south east, inner sep=0pt,xscale=-1] at (axis cs:0,0)
    {\includegraphics[width=22.2mm, height=12.5mm]{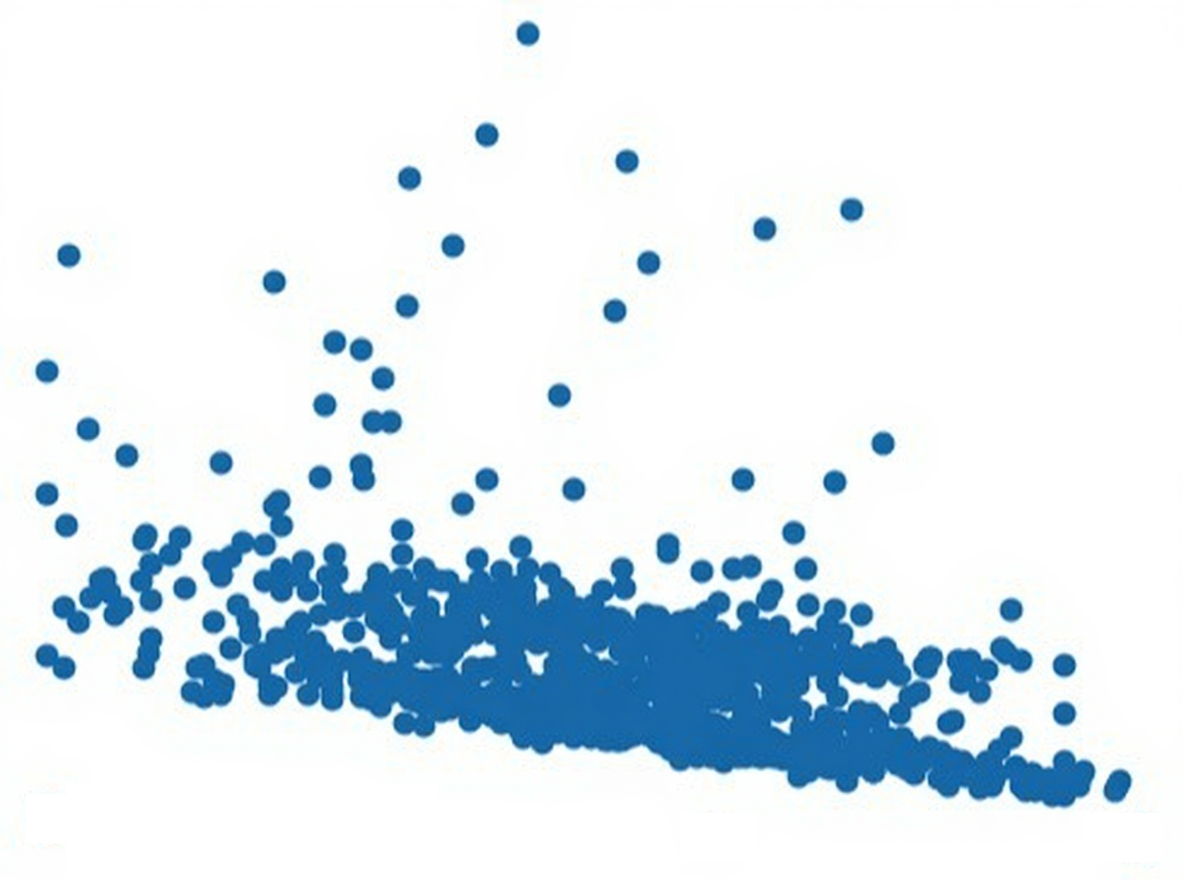}};
% ## 4times
% 3times
\end{axis}
\end{tikzpicture}
\vspace{-2.5mm}
    \caption{\spann, \texttt{nprobe=2048}}
    \label{fig:cache_gains_spann}
\end{subfigure}
\hfill
\begin{subfigure}[b]{0.49\linewidth}
\centering
\begin{tikzpicture}

\begin{axis}[
    xtick=data,
    width=38mm,
height=28mm,
    ymin=0,
    ymax=6,
    clip=false,
    axis y line*=none,
    axis x line*=none,
    tick align=outside,
    ytick={0, 1, 2, 3, 4, 5, 6},
    yticklabels={0, 1, 2, 3, 4, 5, 6},
    xlabel=Num. roundtrips,
    xlabel style={yshift = 1.5ex},
    label style={font=\scriptsize},
        ylabel style={yshift=-1ex,xshift=-0.5ex, font=\scriptsize},
    xmin = 0,
    xmax = 45,
    xtick={0, 10, 20, 30, 40},
    xticklabels={0, 10, 20, 30, 40},
    x tick label style={yshift=0.5ex},
    tick label style={font=\scriptsize},
    legend style={
        at={(-0.2,1.1)},anchor=south west,column sep=2pt,
        draw=black,fill=white,
        inner ysep=0.1pt,
        /tikz/every even column/.append style={column sep=5pt},
        font=\footnotesize
    },
    legend cell align={left},
    legend columns=5,
    ylabel={Query latency (s)},
    ymajorgrids,
    every axis plot/.append style={thick}
    % legend image code/.code={%
    % \draw[#1, draw=none] (0cm,-0.1cm) rectangle (0.6cm,0.1cm);}
]

% \addplot[line width=1pt, GreenColor,mark=*] coordinates
% {(1, 51.2) (2, 47.7) (3, 50.5) (4, 62.2) (5, 63.0)(6, 75.2)};
% Read
\node[anchor=south west, inner sep=0pt] at (axis cs:0,0)
    {\includegraphics[width=22.2mm, height=12.5mm]{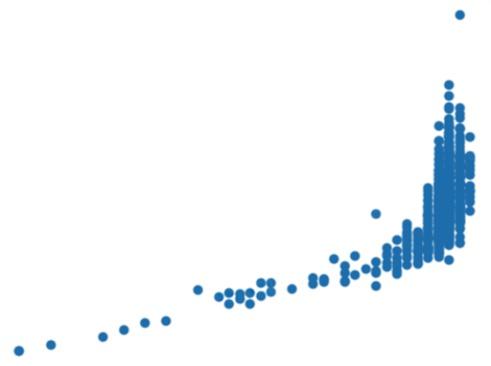}};
% ## 4times
% 3times
\end{axis}
\end{tikzpicture}
\vspace{-2.5mm}
\caption{\diskann, \texttt{search\_len=640}}
\label{fig:cache_gains_diskann}
\end{subfigure}
\vspace{-1mm}
\caption{\spann and \diskann queries on \gist with 4GB SLRU cache; setup described in \cref{sec:exp_setup}. Queries' latencies are correlated with data read and number of roundtrips, respectively, which cache hits can help reduce.}
\label{fig:cache_gains}
\vspace{-3mm}
\end{figure}

\subsection{Caching Benefits Vector Search}
\label{sec:caching_background}
Works and surveys have found vector search workloads in production to exhibit \textit{commonality} and \textit{stability} patterns in production scenarios~\cite{mohoney2023high, mageirakos2025cracking, mohoney2024incremental, zhou2025govector}, where the access patterns of posting lists of cluster-based indexes~\cite{mohoney2024incremental} and nodes of graph-based indexes~\cite{zhou2025govector} are both long-tailed and stable over time.
These patterns have lead to effective caching solutions for related analytical tasks~\cite{schmidt2024predicate, jindal2018computation}, and has motivated recent studies of caching for vector search~\cite{jeong2025cagr, zhou2025govector}.

% \paragraph{Caching of Index Metadata Structures}
% Like traditional inverted indexes holding the keyword list in memory~\cite{wu2013ginix}, Many existing disk-based vector indexes statically cache metadata structures accessed by every query in memory for the duration of query serving: examples include SPANN's BKT tree for finding relevant centroids~\cite{chen2021spann}, DiskANN's Product-Quantized dataset for distance computations during graph traversal~\cite{jayaram2019diskann}, and IVF-PQ's coarse centroid list and codebook~\cite{ivfpq}.  TurboPuffer~\cite{turbopufferarchitecture} uses a two-tier RAM/LRU cache to hold frequently accessed SPFresh indexes' BKT trees.

\paragraph{Caching Index Segments}
Existing caching for vector search works using well-established caching strategies: for example, GoVector~\cite{zhou2025govector} and CrackingIVF~\cite{mageirakos2025cracking} uses an LRU cache to hold frequently visited nodes of the DiskANN index and IVF posting lists, respectively. Like many prior caching works, these works aim to maximize the cache hit rate of query workloads, and have shown via experimentation that higher cache hit rate within a query (e.g., more nodes fetched from cache) \textit{in general} leads to lower query latency~\cite{zhou2025govector}.

\paragraph{Caching to Address Cloud Bottlenecks}
Accordingly, we preliminarily study of the effectiveness of these caching strategies for cloud-native vector search, by performing a cold run of \gist's 1,000 queries sequentially under a high-recall (0.995), hence I/O constrained, scenario with \spann (\cref{fig:cache_gains_spann}) and \diskann (\cref{fig:cache_gains_diskann}), using a 4GB SLRU cache (details in \cref{sec:exp_setup}) to hold hot posting lists and index blocks, respectively. We expectedly observe that caching can benefit on-cloud search: cache hits for \spann and \diskann queries may reduce the data read and roundtrip count of individual queries, which, dominating the indexes' cost models under high-recall settings (\cref{sec:cloud_index_cost}), would accordingly decrease query latency. However, these plots also contain peculiar patterns: while the distribution of data read per query is more uniform for \spann (\cref{fig:cache_gains_spann}), the majority of queries receive few benefits from caching for \diskann (\cref{fig:cache_gains_diskann}), which the next section will explore (\cref{sec:caching_benefit}).

% which leaves an important question to be studied: \textit{how much does query latency and throughput actually benefit from cache hits?} We will present some preliminary analysis to set up for our experimentation in \cref{sec:caching_benefit}.
% \begin{enumerate}
%     \item Caching has been extensively explored for analytical tasks:\begin{enumerate}
%         \item OLAP: intermediate tables and/or results are cached
%         \item KV stores: key-value pairs are cached
%         \item Many different eviction policies
%     \end{enumerate}
%     \item Caching items for vector search\begin{enumerate}
%         \item Like with on-disk indexing, current disk-based indexes maintain a small amout of (meta)data in the cache
%         \item DiskANN: dataset quantized with PQ
%         \item SPANN: BKT tree
%     \end{enumerate}
%     \item Turbopuffer's cache for vector search\begin{enumerate}
%         \item Cache + SPANN; hot posting lists are held in memory
%         \item Two-tier cache: Memory + SSD
%     \end{enumerate}
% \end{enumerate}

\input{sections/plots/diskann_cache_teaser}
\subsection{Index Optimizations May Harm Caching}
\label{sec:caching_benefit}
Optimizations for on-disk indexes such as \spann's vector replication and \diskann's multi-width beam search (\cref{sec:cloud_index_improve}) are largely orthogonal to caching for vector search. Yet, the two may interact in surprising ways, which this section will overview.

% This section studies the extent which graph and cluster-based vector indexes benefit from caching, in particular on remote storage. Without loss of generality, we will be using the Krypton~\cite{chen2023krypton} in-memory cache with the scan-resistant LRU eviction policy to perform some preliminary studies, following indexing setups described in \cref{sec:exp_setup}.

\paragraph{Multi-Width Beam Search Decreases Caching Gains} We expand upon \cref{fig:cache_gains_diskann} and study the cache hit rate metrics of \gist's workload run with \diskann in \cref{fig:diskann_cache_teaser}.
We can observe a noticeable non-linear relationship between query-wide hitrate (i.e., \% of nodes/index blocks that were fetched from the cache during traversal) and number of performed roundtrips: the number of roundtrips to remote storage only significantly drops when cache hit rate is very high (>0.9). This is because \diskann's multi-width beam search ($W=16$ in this case) \textit{actively diminish} caching gains under non-IOPS saturated scenarios:\footnote{we explore this scenario in \cref{sec:exp_caching}.} \textit{all} $W$ candidates of an expansion round must be cached to save a roundtrip to remote storage (\cref{fig:diskann_cache_teaser_expansion}), which, the larger $W$ is, the less likely this would happen and the fewer gains under the same cache size (verified in \cref{sec:exp_caching}). Consequently, we observe the same non-linear relationship between query-wide hitrate and latency (\cref{fig:diskann_cache_teaser_latency}). 

\paragraph{(RQ3) How to utilize caching?} 
Other negative examples between vector index optimizations and caching exist, for example, \spann's vector replication increasing posting list size and consequently decreasing cache hit rate under the same cache size. 
Hence, it is clear that optimal index parameterization for cloud-native vector search depends on available local cache size, a specific examplle being whether large \diskann $W$ values should still be used \textit{despite} reductions in caching gains.
This leads us to study the question \textit{How should caching be utilized for on-cloud search, and how would caching also affect optimal index parameterization?} in \cref{sec:exp_caching}.

\section{Experiments}
\label{sec:experiments}
This section empirically studies our proposed research questions on cloud-native vector search. We structure our study as follows:
% Bottleneck study of on-remote storage querying for different index forms (e.g., numbers like mean I/O, flame graphs)
\begin{enumerate}
    \item \textbf{Factors Affecting Cloud Indexes (\cref{sec:exp_e2e}):} We compare performance and bottlenecks of cluster and graph indexes on datasets and workloads with varying data dimension, concurrencies, etc., to find which scenarios each index class is performant on. 
    \item \textbf{Index Design for Cloud Storage (\cref{sec:exp_parameterization}):} We observe how index classes' build and query parameterizations affect performance trade-offs on cloud storage---e.g., data read, computations, to provide recommendations on how to tune the indexes according to specific workload and environment characteristics.
    \item \textbf{On-Cloud Index-Cache Integration (\cref{sec:exp_caching}):} We explore gains achieved by integrating existing caching strategies applicable to caching index segments---e.g., how cache hits benefit query performance, to gain insights on how to coordinate index design and cache characteristics for maximum gains.
    % \item \textbf{Adapting indexes to remote storage:} Given the bottlenecks of each index class on remote storage, how can we adjust the indexes (e.g., by tuning index parameterization) to improve their performance on remote storage? (\cref{sec:exp_parameterization})
    % \item \textbf{Caching for on-remote storage vector search:} How cache-friendly are the index classes? How do the performance benefits change with available cache size, and how can we adjust indexes in each index class to make them more cache-friendly? (\cref{sec:exp_caching})
    % \item \textbf{Cache-aware indexing:} Which adjustments can we make to indexes and/or caching policies to improve the performance gains from caching? (\cref{sec:exp_detailed})
    % \item \textbf{Case studies:} We give recommendations on the index class, parameterization, and querying setup to use with various emerging real-world scenarios calling for on-remote storage vector search. (\cref{sec:exp_casestudy})
\end{enumerate}
\begin{table}[t]
\caption{Summary of Datasets for Evaluation.}
\footnotesize
\vspace{-1mm}
\addtolength{\tabcolsep}{-1pt} 
\begin{tabular}{l rr l r l}
\toprule
\textbf{Dataset} & \textbf{\# Vectors} & \textbf{Dim}  & \textbf{Data Type} & \textbf{\# Queries} & \textbf{Modality} \\
 \midrule
\gist~\cite{gist}& 1000000 & 960 & FLOAT32 & 1000 & Image\\
\deep~\cite{babenko2016efficient} & 10000000 & 96 & INT8 &  10000 & Image\\
\msspace~\cite{uqv}& 10000000 & 100 & FLOAT32 & 30000 & Document\\
\bigann~\cite{bigann1b} & 100000000 & 128 & INT8 &  10000 & Image\\

\bottomrule
\end{tabular}
\vspace{-3mm}
\addtolength{\tabcolsep}{1pt}
\label{tbl:workload}
\end{table}

\subsection{Experiment setup}
\label{sec:exp_setup}
\paragraph{Datasets} We use 4 experiment datasets (\cref{tbl:workload}): \gist and \deep are from the ANN benchmark~\cite{annbench}; \msspace and \bigann are from the Neurips'21 BigANN benchmark~\cite{bigann1b}. We search for $k=10$ nearest neighbors for each query in each dataset.

\begin{table}[t]
\caption{Summary of Default Indexing/Search Parameters for Evaluation on each Dataset Unless Otherwise Stated.}
\footnotesize
\vspace{-1mm}
% \addtolength{\tabcolsep}{-1pt} 
\begin{tabular}{|l|rrr|rr|}
\toprule
 \multicolumn{1}{|l|}{\textbf{Index}} & \multicolumn{3}{c|}{\diskann} & \multicolumn{2}{c|}{\spann} \\
  \midrule
\textbf{Dataset} & \textbf{PQ dim.} & \textbf{M} & \textbf{Beamwidth}& \textbf{centroid\%}  & \textbf{replica\#} \\
 \midrule
\gist~\cite{gist}& 112 & 64 & 16 & 16 & 8 \\
\deep~\cite{babenko2016efficient} & 48 & 64 & 16 & 12 & 8 \\
\msspace~\cite{uqv}& 112  & 64 & 16 & 12 & 8 \\
\bigann~\cite{bigann1b} & 112 & 64 & 16 & 12 & 8 \\

\bottomrule
\end{tabular}
\vspace{-3mm}
% \addtolength{\tabcolsep}{1pt}
\label{tbl:parameters}
\end{table}

\paragraph{Methods} We study the following indexes as the representatives from their respective index classes (parameterizations in \cref{tbl:parameters}):
\begin{itemize}
    \item \textbf{DiskANN~\cite{jayaram2019diskann}} is the state-of-the-art on-disk graph-based index. Unless stated (e.g., \cref{sec:exp_parameterization}), we use $L=128, R=64, W=16$, $sector\_len=4KB$, and $QD=max(\frac{dim}{8}, 48)$ across all datasets.
    \item \textbf{SPANN~\cite{chen2021spann}} is the state-of-the-art on-disk cluster-based index. Unless stated (e.g., \cref{sec:exp_parameterization}), we use the more performant out of $centroid\%=[12,16]$ and $num\_replica=8$ across all datasets.
\end{itemize}
\paragraph{System setup} We perform all experiments on a ByteDance ecs.s2-c1m4.14xlarge machine with 56 vCPUs, 224GB RAM and 100GB NIC bandwidth. We use the Volcano Engine TOS~\cite{volcenginetos} for our remote storage. We build our indexes on local disk, which we then upload to remote storage. We utilize external network to access the remote storage, which has a download network bandwidth to our machine of 5Gbps, a GET request limit of 20,000QPS, and a p50 read latency of 31,000 microseconds (\cref{tbl:env_diff}).
For caching experiments in \cref{sec:exp_caching}, we use a ByteDance in-house in-memory cache which features asynchronous data reads and writes to and from the cache~\cite{chen2023krypton}. We cache posting lists (for \spann) and (4KB) index blocks (for \diskann) following a scan-resistant LRU eviction policy~\cite{megiddo2004outperforming}.
% For experiments involving local disk, we use an NVMe SSD with a read speed of 12GB/s and read latency of 20 microseconds.

\paragraph{Query Serving} We load the relevant index metadata into memory, i.e., BKT trees for \spann and PQ dataset for \diskann prior to serving each query workload on each dataset following current local caching implementations in cloud-native vector search (\cref{sec:caching}). Then we sweep $search\_len\in[10, 1280]$ in power of 2 increments for DiskANN and $nprobe\in[8, 16384]$ in power of 2 increments for SPANN to generate QPS-recall curves, early stopping if the current parameterization value achieves a recall $>0.995$. We also sweep number of concurrently served queries from 1 to 64 for certain scenarios, which we controll via queueing.

\paragraph{Measurement} At the workload level, we measure the \textcircled{1} \textit{throughput} as queries-per-second (QPS) of query serving, \textcircled{2} \textit{latency} (at different percentiles) as time to serve each query, and \textcircled{3} average network bandwidth,\footnote{Computed as total data read/real elapsed workload time.}. For index-specific metrics, we measure \textcircled{4} \textit{number of neighbor expansions} performed for each DiskANN query and the \textcircled{5} \textit{number of visited posting lists} for each SPANN query. For query-specific metrics, we measure \textcircled{6} per-request \textit{mean I/O latency} of individual \spann posting lists and \diskann per-expansion node batches,\footnote{Implementation-wise, \spann issues separate requests for posting lists~\cite{sptaggithub}, while \diskann batches the $W$ requests of each expansion~\cite{diskanngithub}. On the remote storage side, \diskann's $W$ batched requests will still count as $W$ IOs for IOPS throttling~\cite{volcenginetos}.} and \textcircled{7} \textit{cache hit rate} for settings with a cache.

\paragraph{Reproducibility} Our benchmarking scripts and datasets are open-sourced in our Github Repository~\cite{ourrepo}.

\input{sections/plots/exp_diskann_vs_spann_all_datasets}
\subsection{What Index For What Scenario?}
\label{sec:exp_e2e}
This section studies the comparison between DiskANN and SPANN indexes on remote storage. We build the two indexes for each dataset following procedures described in \cref{sec:exp_setup}, then study the metrics and performance of each index on different query workloads.

We report results in \cref{fig:exp_spann_vs_diskann}. We can observe that \diskann outperforms \spann in high-concurrency and/or recall scenarios on all datasets, matching observations in prior studies for on-disk indexing~\cite{cheng2024characterizing, renstorage}. Focusing on results on \gist (\cref{fig:exp_spann_vs_diskann_gist}), we can see that the cutoff is at $\sim$0.8 recall with 16 concurrent queries (\cref{fig:exp_spann_vs_diskann_gist_concurrency16}); \spann outperforms \diskann at all recalls at lower concurrencies (\cref{fig:exp_spann_vs_diskann_gist_concurrency1}, \cref{fig:exp_spann_vs_diskann_gist_concurrency4}), and vice versa at higher concurrencies (\cref{fig:exp_spann_vs_diskann_gist_concurrency64}). 
\begin{figure}[t]\captionsetup[subfigure]{font=footnotesize}
\pgfplotsset{scaled y ticks=false}
\centering
\begin{subfigure}[b]{0.32\linewidth}
\begin{tikzpicture}

\begin{axis}[
    xtick=data,
    width=32mm,
    height=28mm,
    ymin=0.1,
    ymax=1000,
    log origin = infty,
    ymode = log,
    axis y line*=none,
    axis x line*=none,
    ytick={0.1, 1, 10, 100, 1000},
    yticklabels={$10^{-1}$, $10^0$, $10^1$, $10^2$,$10^3$},
    xlabel=Recall,
    xlabel style={yshift = 1.5ex},
    label style={font=\scriptsize},
    ylabel style={yshift=-1ex,xshift=-0.5ex, font=\scriptsize},
    xmin = 0.6,
    xmax = 1,
    xtick = {0.6, 0.7, 0.8, 0.9, 1.0},
    xticklabels = {0.6, 0.7, 0.8, 0.9, 1.0},
    tick label style={font=\scriptsize},
    x tick label style={yshift=0.5ex},
    legend style={
        at={(0,1.09)},anchor=south west,column sep=0pt,
        legend image post style={xscale=0.6},
        row sep = -0.4pt,
        draw=black,fill=white,
        inner ysep=0.1pt,
        /tikz/every even column/.append style={column sep=2pt},
        font=\footnotesize
    },
    legend cell align={left},
    legend columns=4,
    ylabel={Data Read (MB)},
    ymajorgrids,
    every axis plot/.append style={thick}
    % legend image code/.code={%
    % \draw[#1, draw=none] (0cm,-0.1cm) rectangle (0.6cm,0.1cm);}
]

% \addplot[line width=1pt, GreenColor,mark=*] coordinates
% {(1, 51.2) (2, 47.7) (3, 50.5) (4, 62.2) (5, 63.0)(6, 75.2)};
% Read
\addplot[line width=1pt, GreenColor, densely dotted]
table[x=x,y=y] {
x y
0.6525  1.331
0.7245  2.578
0.8033  4.979
0.8767  9.708
0.9334  18.827
0.9690  36.447
0.9849  70.349
0.9925  134.396
0.9963  256.214
};
\addlegendentry{\spann}
\addplot[line width=1pt, HeuristicColor, densely dashed]
table[x=x,y=y] {
x y
0.5812  0.157
0.7563  0.275
0.8474  0.346
0.9177  0.477
0.9641  0.777
0.9885  1.411
0.9955  2.703
};
\addlegendentry{\diskann}
% ## 4times

\end{axis}
\end{tikzpicture}
\vspace{-6.5mm}
\caption{Data Read per Query}
\label{fig:exp_index_metrics_dataread}
\end{subfigure}
\begin{subfigure}[b]{0.32\linewidth}
\begin{tikzpicture}

\begin{axis}[
    xtick=data,
    width=32mm,
    height=28mm,
    ymin=0,
    ymax=50,
    axis y line*=none,
    axis x line*=none,
    ytick={0, 10, 20, 30, 40, 50},
    ytick={0, 10, 20, 30, 40, 50},
    xlabel=Recall,
    xlabel style={yshift = 1.5ex},
    label style={font=\scriptsize},
    ylabel style={yshift=-1ex,xshift=-0.5ex, font=\scriptsize},
    xmin = 0.6,
    xmax = 1,
    xtick = {0.6, 0.7, 0.8, 0.9, 1.0},
    xticklabels = {0.6, 0.7, 0.8, 0.9, 1.0},
    tick label style={font=\scriptsize},
    x tick label style={yshift=0.5ex},
    legend style={
        at={(-0.4,1.09)},anchor=south west,column sep=0pt,
        legend image post style={xscale=0.6},
        row sep = -0.4pt,
        draw=black,fill=white,
        inner ysep=0.1pt,
        /tikz/every even column/.append style={column sep=2pt},
        font=\footnotesize
    },
    legend cell align={left},
    legend columns=4,
    ylabel={Num. Roundtrips},
    ymajorgrids,
    every axis plot/.append style={thick}
    % legend image code/.code={%
    % \draw[#1, draw=none] (0cm,-0.1cm) rectangle (0.6cm,0.1cm);}
]

% \addplot[line width=1pt, GreenColor,mark=*] coordinates
% {(1, 51.2) (2, 47.7) (3, 50.5) (4, 62.2) (5, 63.0)(6, 75.2)};
% Read
\addplot[line width=1pt, GreenColor, densely dotted]
table[x=x,y=y] {
x y
0.6525  1
0.7245  1
0.8033  1
0.8767  1
0.9334  1
0.9690  1
0.9849  1
0.9925  1
0.9963  1
};
\addplot[line width=1pt, HeuristicColor, densely dashed]
table[x=x,y=y] {
x y
0.5812  7
0.7563  7
0.8474  8
0.9177  10
0.9641  14
0.9885  24
0.9955  43
};

% ## 4times

\end{axis}
\end{tikzpicture}
\vspace{-2.5mm}
\caption{Roundtrips per Query}
\label{fig:exp_index_metrics_roundtrips}
\end{subfigure}
\begin{subfigure}[b]{0.32\linewidth}
\begin{tikzpicture}

\begin{axis}[
    xtick=data,
    width=32mm,
    height=28mm,
    ymin=1,
    ymax=10000,
    log origin = infty,
    ymode = log,
    axis y line*=none,
    axis x line*=none,
    ytick={1, 10, 100, 1000, 10000},
    yticklabels={$10^0$, $10^1$, $10^2$,$10^3$, $10^4$},
    xlabel=Recall,
    xlabel style={yshift = 1.5ex},
    label style={font=\scriptsize},
    ylabel style={yshift=-1ex,xshift=-0.5ex, font=\scriptsize},
    xmin = 0.6,
    xmax = 1,
    xtick = {0.6, 0.7, 0.8, 0.9, 1.0},
    xticklabels = {0.6, 0.7, 0.8, 0.9, 1.0},
    tick label style={font=\scriptsize},
    x tick label style={yshift=0.5ex},
    legend style={
        at={(-0.4,1.09)},anchor=south west,column sep=0pt,
        legend image post style={xscale=0.6},
        row sep = -0.4pt,
        draw=black,fill=white,
        inner ysep=0.1pt,
        /tikz/every even column/.append style={column sep=2pt},
        font=\footnotesize
    },
    legend cell align={left},
    legend columns=4,
    ylabel={Num. requests},
    ymajorgrids,
    every axis plot/.append style={thick}
    % legend image code/.code={%
    % \draw[#1, draw=none] (0cm,-0.1cm) rectangle (0.6cm,0.1cm);}
]

% \addplot[line width=1pt, GreenColor,mark=*] coordinates
% {(1, 51.2) (2, 47.7) (3, 50.5) (4, 62.2) (5, 63.0)(6, 75.2)};
% Read
\addplot[line width=1pt, GreenColor, densely dotted]
table[x=x,y=y] {
x y
0.6525  8
0.7245  16
0.8033  32
0.8767  64
0.9334  128
0.9690  256
0.9849  512
0.9925  1024
0.9963  2048
};
\addplot[line width=1pt, HeuristicColor, densely dashed]
table[x=x,y=y] {
x y
0.5812  38.45
0.7563  67.19
0.8474  84.58
0.9177  116.56
0.9641  189.85
0.9885  344.5
0.9955  660.16
};

% ## 4times

\end{axis}
\end{tikzpicture}
\vspace{-2.5mm}
\caption{Requests per Query}
\label{fig:exp_index_metrics_requests}
\end{subfigure}
\vspace{-2mm}
\caption{Query metrics of \spann and \diskann on \gist. \spann reads significantly more data per query versus \diskann, but \diskann's requests are spread across roundtrips due to dependencies within graph traversal (\cref{sec:cloud_index_cost}).
}
\vspace{-3mm}
\label{fig:exp_index_metrics}
\end{figure}
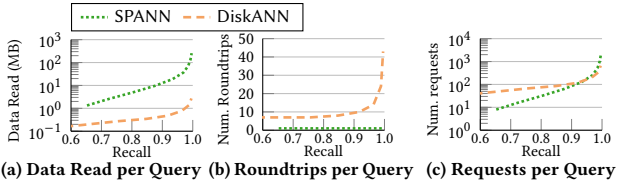
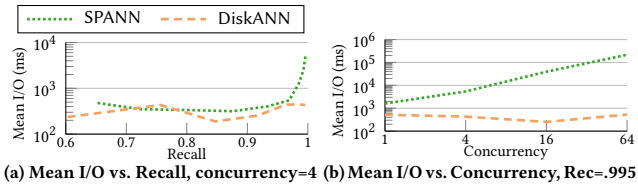
\begin{figure}[t]\captionsetup[subfigure]{font=footnotesize}
\pgfplotsset{scaled y ticks=false}
\centering

\begin{subfigure}[b]{0.49\linewidth}
\begin{tikzpicture}

\begin{axis}[
    xtick=data,
    width=48mm,
height=28mm,
    ymin=10,
    ymax=1000,
    log origin = infty,
    ymode = log,
    axis y line*=none,
    axis x line*=none,
    ytick={10, 100, 1000},
    yticklabels={$10^2$, $10^3$, $10^4$},
    xlabel=Recall,
    xlabel style={yshift = 1.5ex},
    label style={font=\scriptsize},
        ylabel style={yshift=-1ex,xshift=-0.5ex, font=\scriptsize},
    xmin = 0.6,
    xmax = 1,
    xtick = {0.6, 0.7, 0.8, 0.9, 1},
    xticklabels = {0.6, 0.7, 0.8, 0.9, 1},
    x tick label style={yshift=0.5ex},
    tick label style={font=\scriptsize},
    legend style={
        at={(-0.2,1.1)},anchor=south west,column sep=2pt,
        draw=black,fill=white,
        inner ysep=0.1pt,
        /tikz/every even column/.append style={column sep=5pt},
        font=\footnotesize
    },
    legend cell align={left},
    legend columns=5,
    ylabel={Mean I/O (ms)},
    ymajorgrids,
    every axis plot/.append style={thick}
    % legend image code/.code={%
    % \draw[#1, draw=none] (0cm,-0.1cm) rectangle (0.6cm,0.1cm);}
]

% \addplot[line width=1pt, GreenColor,mark=*] coordinates
% {(1, 51.2) (2, 47.7) (3, 50.5) (4, 62.2) (5, 63.0)(6, 75.2)};
% Read
\addplot[line width=1pt, GreenColor, densely dotted]
table[x=x,y=y] {
x y
0.6525  47.825
0.7245  35.049
0.8033  33.503
0.8767  31.622
0.9334  40.427
0.9690  54.462
0.9849  126.059
0.9925  245.477
0.9963  542.075
};
\addlegendentry{\spann}
\addplot[line width=1pt, HeuristicColor, densely dashed]
table[x=x,y=y] {
x y
0.5812  21.66828571
0.7563  43.13525581
0.8474  18.897875
0.9177  25.5059
0.9641  44.20771429
0.9885  44.21226087
0.9955  43.13525581
};
\addlegendentry{\diskann}
% ## 4times
% 3times
a\end{axis}
\end{tikzpicture}
\vspace{-6.5mm}
\caption{Mean I/O vs. Recall, concurrency=4}
\label{fig:exp_meanio_gist_recall}
\end{subfigure}
\hfill
\begin{subfigure}[b]{0.49\linewidth}
\begin{tikzpicture}

\begin{axis}[
    xtick=data,
    width=48mm,
height=28mm,
    ymin=10,
    ymax=100000,
    log origin = infty,
    ymode = log,
    axis y line*=none,
    axis x line*=none,
    ytick={10, 100, 1000, 10000, 100000},
    yticklabels={$10^2$, $10^3$, $10^4$, $10^5$, $10^6$},
    xlabel=Concurrency,
    xlabel style={yshift = 1.5ex},
    label style={font=\scriptsize},
        ylabel style={yshift=-1ex,xshift=-0.5ex, font=\scriptsize},
    xmin = 0,
    xmax = 3,
    xtick = {0, 1, 2, 3},
    xticklabels = {1, 4, 16, 64},
    x tick label style={yshift=0.5ex},
    tick label style={font=\scriptsize},
    legend style={
        at={(-0.2,1.1)},anchor=south west,column sep=2pt,
        draw=black,fill=white,
        inner ysep=0.1pt,
        /tikz/every even column/.append style={column sep=5pt},
        font=\footnotesize
    },
    legend cell align={left},
    legend columns=5,
    ylabel={Mean I/O (ms)},
    ymajorgrids,
    every axis plot/.append style={thick}
    % legend image code/.code={%
    % \draw[#1, draw=none] (0cm,-0.1cm) rectangle (0.6cm,0.1cm);}
]

% \addplot[line width=1pt, GreenColor,mark=*] coordinates
% {(1, 51.2) (2, 47.7) (3, 50.5) (4, 62.2) (5, 63.0)(6, 75.2)};
% Read
\addplot[line width=1pt, GreenColor, densely dotted]
table[x=x,y=y] {
x y
0 161.423
1 542.075
2 3933.236
3 21641.523
};
\addplot[line width=1pt, HeuristicColor, densely dashed]
table[x=x,y=y] {
x y
0 51.9976744
1 43.13525581
2 25.29418605
3 52.94727907
};
% ## 4times
% 3times
\end{axis}
\end{tikzpicture}
\vspace{-6.5mm}
\caption{Mean I/O vs. Concurrency, Rec=.995}
\label{fig:exp_meanio_gist_concurrency}
\end{subfigure}
\vspace{-2mm}
\caption{Mean I/O of on-disk vs. on-remote storage querying of \spann on \gist. \spann exhibits notable I/O congestion at both high recalls and concurrencies.
% These costs are negligible on notebooks with up to 2000 cell executions.
}
\vspace{-3mm}
\label{fig:exp_meanio_gist}
\end{figure}

% \spann's I/O bandwidth for on-cloud querying exhibits bottlenecking at both high-recall and high-concurrency scenarios, which both contend for I/O bandwidth due to the index's concurrent queuing---both inter and intra-query (\textcolor{red}{cref})---of (often large) posting lists (studied shortly in \cref{sec:exp_bottleneck}). \diskann's performance is significantly less sensitive to I/O bandwdith, exhibiting consistent bandwidth utilization at different recall levels, and thanks to the significantly smaller request sizes---4KB blocks (\textcolor{red}{cref}) vs. \spann's large posting lists, also exhibit better scaling to high concurrencies, using only 11.7\% of the available I/O bandwidth (77MB/s) at 64 concurrent queries with 0.995 recall.

The detailed query metrics in \cref{fig:exp_index_metrics} and mean I/O latencies in \cref{fig:exp_meanio_gist} hint at the reasons: each \spann query reads a large amount of data, from 2.5MB at 0.7 recall to 256MB at 0.995 recall, which is 9.39$\times$ and 94.8$\times$ more than \diskann at the same recalls, respectively (\cref{fig:exp_index_metrics_dataread}).
Hence, \spann exhibits significant I/O congestion at high recalls (\cref{fig:exp_meanio_gist_recall}) and concurrencies (\cref{fig:exp_meanio_gist_concurrency}) as these two settings both contend for I/O bandwidth (\cref{sec:background_characteristics}) due to the index's concurrent queuing---both inter and intra-query (\cref{sec:cluster_index_cost})---of posting lists: at 0.995 recall and 64 concurrent queries, \spann's posting list mean I/O latency is \textit{21.6 seconds}, and the bandwidth is only enough to serve \textit{2.5} \spann queries per second at this recall.
Conversely, \diskann's exhibits significantly less I/O congestion---its batched node expansion requests having 416$\times$ lower mean I/O latency versus \spann at 0.995 recall and 64 concurrent queries, due to its iterative traversal spreading out its (already low) data requests to exhibit consistent bandwidth utilization at different recalls (\cref{fig:teaser_io_usage_diskann}).

Yet, the disadvantage of \diskann's iterative traversal becomes apparent under lower conurrency settings: it makes multiple roundtrips per query due to its iterative traversal, from 7 at 0.7 recall to 43 at 0.995 recall (\cref{fig:exp_index_metrics_roundtrips}). \diskann queries have a latency of 285 ms when sequentially running queries at 0.7 recall (\cref{fig:exp_spann_vs_diskann_gist_concurrency1}), where if we follow the read latencies observed in \cref{tbl:env_diff}, suggests that on average \textit{217 ms} was spent on communicating with remote storage alone. Inability to fully utilize I/O bandwidth for individual queries unlike \spann's concurrent fetching of posting lists leads to \diskann having 5.71$\times$ higher versus \spann at this setting.

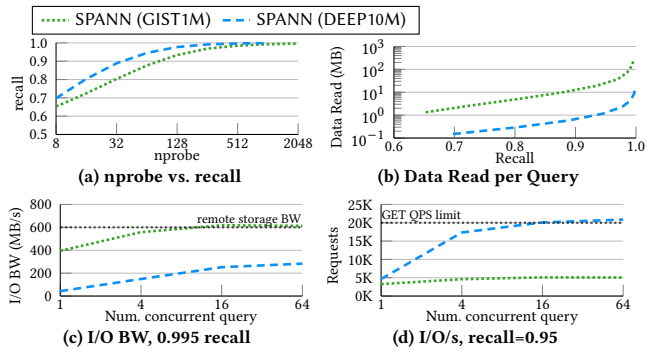
\begin{figure}[t]\captionsetup[subfigure]{font=footnotesize}
\pgfplotsset{scaled y ticks=false}
\centering
\begin{subfigure}[b]{0.49\linewidth}
\begin{tikzpicture}

\begin{axis}[
    xtick=data,
    width=48mm,
    height=28mm,
    ymin=0.5,
    ymax=1,
    axis y line*=none,
    axis x line*=none,
    ytick={0.5, 0.6, 0.7, 0.8, 0.9, 1.0},
    yticklabels={0.5, 0.6, 0.7, 0.8, 0.9, 1.0},
    xlabel=nprobe,
    xlabel style={yshift = 1.5ex},
    label style={font=\scriptsize},
    ylabel style={yshift=-1ex,xshift=-0.5ex, font=\scriptsize},
    xmin = 0,
    xmax = 8,
    xtick={0, 2, 4, 6, 8},
    xticklabels={8, 32, 128, 512, 2048},
    tick label style={font=\scriptsize},
    x tick label style={yshift=0.5ex},
    legend style={
        at={(-0.1,1.09)},anchor=south west,column sep=0pt,
        legend image post style={xscale=0.6},
        row sep = -0.4pt,
        draw=black,fill=white,
        inner ysep=0.1pt,
        /tikz/every even column/.append style={column sep=2pt},
        font=\footnotesize
    },
    legend cell align={left},
    legend columns=4,
    ylabel={recall},
    ymajorgrids,
    every axis plot/.append style={thick}
    % legend image code/.code={%
    % \draw[#1, draw=none] (0cm,-0.1cm) rectangle (0.6cm,0.1cm);}
]

% \addplot[line width=1pt, GreenColor,mark=*] coordinates
% {(1, 51.2) (2, 47.7) (3, 50.5) (4, 62.2) (5, 63.0)(6, 75.2)};
% Read
\addplot[line width=1pt, GreenColor, densely dotted]
table[x=x,y=y] {
x y
0 0.6525 
1 0.7245 
2 0.8033 
3 0.8767 
4 0.9334 
5 0.9690 
6 0.9849 
7 0.9925 
8 0.9963 
};
\addlegendentry{\spann (\gist)}
\addplot[line width=1pt, AcornOneColor, densely dashed]
table[x=x,y=y] {
x y
0 0.6977
1 0.8035
2 0.8886
3 0.9453
4 0.9773
5 0.9914
6 0.997
7 0.9991
};
\addlegendentry{\spann (\deep)}
% ## 4times

\end{axis}
\end{tikzpicture}
\vspace{-6.5mm}
\caption{nprobe vs. recall}
\label{fig:exp_spann_deep_recall}
\end{subfigure}
\begin{subfigure}[b]{0.49\linewidth}
\begin{tikzpicture}

\begin{axis}[
    xtick=data,
    width=48mm,
    height=28mm,
    ymin=0.1,
    ymax=1000,
    log origin = infty,
    ymode = log,
    axis y line*=none,
    axis x line*=none,
    ytick={0.1, 1, 10, 100, 1000},
    yticklabels={$10^{-1}$, $10^0$, $10^1$, $10^2$,$10^3$},
    xlabel=Recall,
    xlabel style={yshift = 1.5ex},
    label style={font=\scriptsize},
    ylabel style={yshift=-1ex,xshift=-0.5ex, font=\scriptsize},
    xmin = 0.6,
    xmax = 1,
    xtick = {0.6, 0.7, 0.8, 0.9, 1.0},
    xticklabels = {0.6, 0.7, 0.8, 0.9, 1.0},
    tick label style={font=\scriptsize},
    x tick label style={yshift=0.5ex},
    legend style={
        at={(0,1.09)},anchor=south west,column sep=0pt,
        legend image post style={xscale=0.6},
        row sep = -0.4pt,
        draw=black,fill=white,
        inner ysep=0.1pt,
        /tikz/every even column/.append style={column sep=2pt},
        font=\footnotesize
    },
    legend cell align={left},
    legend columns=4,
    ylabel={Data Read (MB)},
    ymajorgrids,
    every axis plot/.append style={thick}
    % legend image code/.code={%
    % \draw[#1, draw=none] (0cm,-0.1cm) rectangle (0.6cm,0.1cm);}
]

% \addplot[line width=1pt, GreenColor,mark=*] coordinates
% {(1, 51.2) (2, 47.7) (3, 50.5) (4, 62.2) (5, 63.0)(6, 75.2)};
% Read
\addplot[line width=1pt, GreenColor, densely dotted]
table[x=x,y=y] {
x y
0.6525  1.331
0.7245  2.578
0.8033  4.979
0.8767  9.708
0.9334  18.827
0.9690  36.447
0.9849  70.349
0.9925  134.396
0.9963  256.214
};
\addplot[line width=1pt, AcornOneColor, densely dashed]
table[x=x,y=y] {
x y
0.6977 0.149
0.8035 0.296 
0.8886 0.586 
0.9453 1.161 
0.9773 2.298
0.9914 4.549
0.997 9.006
0.9991 17.829
};
% ## 4times

\end{axis}
\end{tikzpicture}
\vspace{-6.5mm}
\caption{Data Read per Query}
\label{fig:exp_spann_deep_dataread}
\end{subfigure}
\hfill
\begin{subfigure}[b]{0.49\linewidth}
\begin{tikzpicture}

\begin{axis}[
    xtick=data,
    width=48mm,
    height=28mm,
    ymin=0,
    ymax=800,
    axis y line*=none,
    axis x line*=none,
    ytick={0, 200, 400, 600, 800},
    yticklabels={0, 200, 400, 600, 800},
    xlabel=Num. concurrent query,
    xlabel style={yshift = 1.5ex},
    label style={font=\scriptsize},
    ylabel style={yshift=-1ex,xshift=-0.5ex, font=\scriptsize},
    xmin = 0,
    xmax = 3,
    xtick = {0, 1, 2, 3},
    xticklabels = {1, 4, 16, 64},
    tick label style={font=\scriptsize},
    x tick label style={yshift=0.5ex},
    legend style={
        at={(-0.6,1.09)},anchor=south west,column sep=0pt,
        legend image post style={xscale=0.6},
        row sep = -0.4pt,
        draw=black,fill=white,
        inner ysep=0.1pt,
        /tikz/every even column/.append style={column sep=2pt},
        font=\footnotesize
    },
    legend cell align={left},
    legend columns=4,
    ylabel={I/O BW (MB/s)},
    ymajorgrids,
    every axis plot/.append style={thick}
    % legend image code/.code={%
    % \draw[#1, draw=none] (0cm,-0.1cm) rectangle (0.6cm,0.1cm);}
]

% \addplot[line width=1pt, GreenColor,mark=*] coordinates
% {(1, 51.2) (2, 47.7) (3, 50.5) (4, 62.2) (5, 63.0)(6, 75.2)};
% Read
\addplot[line width=1pt, GreenColor, densely dotted]
table[x=x,y=y] {
x y
0  395
1  558
2  618
3  613
};
\addplot[line width=1pt, AcornOneColor, densely dashed]
table[x=x,y=y] {
x y
0 43
1 149
2 252
3 283
};
% ## 4times

\draw[black, thick, opacity=0.7, densely dotted] (axis cs: 0, 600) -- (axis cs: 3, 600);
\node[anchor=south east, align=right, inner sep = 0.1pt] at (axis cs: 3, 600) {\tiny remote storage BW};
% ## 4times

\end{axis}
\end{tikzpicture}
\vspace{-6.5mm}
\caption{I/O BW, 0.995 recall}
\label{fig:exp_spann_deep_bw}
\end{subfigure}
\begin{subfigure}[b]{0.49\linewidth}
\begin{tikzpicture}

\begin{axis}[
    xtick=data,
    width=48mm,
    height=28mm,
    ymin=0,
    ymax=25,
    axis y line*=none,
    axis x line*=none,
    ytick={0, 5, 10, 15, 20, 25},
    yticklabels={0K, 5K, 10K, 15K, 20K, 25K},
    xlabel=Num. concurrent query,
    xlabel style={yshift = 1.5ex},
    label style={font=\scriptsize},
    ylabel style={yshift=-1ex,xshift=-0.5ex, font=\scriptsize},
    xmin = 0,
    xmax = 3,
    xtick = {0, 1, 2, 3},
    xticklabels = {1, 4, 16, 64},
    tick label style={font=\scriptsize},
    x tick label style={yshift=0.5ex},
    legend style={
        at={(-0.4,1.09)},anchor=south west,column sep=0pt,
        legend image post style={xscale=0.6},
        row sep = -0.4pt,
        draw=black,fill=white,
        inner ysep=0.1pt,
        /tikz/every even column/.append style={column sep=2pt},
        font=\footnotesize
    },
    legend cell align={left},
    legend columns=4,
    ylabel={Requests},
    ymajorgrids,
    every axis plot/.append style={thick}
    % legend image code/.code={%
    % \draw[#1, draw=none] (0cm,-0.1cm) rectangle (0.6cm,0.1cm);}
]

% \addplot[line width=1pt, GreenColor,mark=*] coordinates
% {(1, 51.2) (2, 47.7) (3, 50.5) (4, 62.2) (5, 63.0)(6, 75.2)};
% Read
\addplot[line width=1pt, GreenColor, densely dotted]
table[x=x,y=y] {
x y
0  3.258
1  4.603
2  5.094
3  5.053
};
\addplot[line width=1pt, AcornOneColor, densely dashed]
table[x=x,y=y] {
x y
0 4.646
1 17.344
2 20.099
3 20.864
};
% ## 4times
\draw[black, thick, opacity=0.7, densely dotted] (axis cs: 0, 20) -- (axis cs: 3, 20);
\node[anchor=south west, align=right, inner sep = 0.1pt] at (axis cs: 0, 20) {\tiny GET QPS limit};
\end{axis}
\end{tikzpicture}
\vspace{-6.5mm}
\caption{I/O/s, recall=0.95}
\label{fig:exp_spann_deep_getqps}
\end{subfigure}
\vspace{-1mm}
\caption{\spann on \deep: lower-dimension dataset reduces posting list size and \texttt{nprobe} values required to reach certain recalls. This significantly reduces data read per query and improves QPS at high recalls and concurrencies, but querying may still be bound by remote storage IOPS.
}
\vspace{-3mm}
\label{fig:exp_spann_deep}
\end{figure}

Finally, despite \diskann reading less data per query versus \spann at all recalls, \diskann makes more requests per query at low recalls (\cref{fig:exp_index_metrics_requests}): at 0.7 recall, each \diskann query averages at 38.4 requests while \spann averages at 7.9 requests; \spann overtakes \diskann at 0.995 recall with 2003.6 and 660.0 requests, respectively. This matches conventional wisdom for the indexes' performance for on-disk querying~\cite{guo2025achieving}, where \diskann is outperformed by \spann at low recalls and vice versa, due to SSD performance being comparatively more sensitive to \diskann's large number of random (4KB) reads at low recalls. For cloud-native vector search, requests per query affects when each index will potentially be bottlenecked by IOPS limits, which we can observe in a variety of settings shortly (e.g., \cref{fig:exp_spann_deep_bw}).

\begin{figure}[t]\captionsetup[subfigure]{font=footnotesize}
\pgfplotsset{scaled y ticks=false}
\centering
\begin{subfigure}[b]{0.32\linewidth}
\begin{tikzpicture}

\begin{axis}[
    xtick=data,
    width=32mm,
    height=28mm,
    ymin=0.5,
    ymax=1,
    axis y line*=none,
    axis x line*=none,
    ytick={0.5, 0.6, 0.7, 0.8, 0.9, 1.0},
    yticklabels={0.5, 0.6, 0.7, 0.8, 0.9, 1.0},
    xlabel=search\_len,
    xlabel style={yshift = 1.5ex},
    label style={font=\scriptsize},
    ylabel style={yshift=-1ex,xshift=-0.5ex, font=\scriptsize},
    xmin = 0,
    xmax = 6,
    xtick={0, 2, 4, 6},
    xticklabels={10, 40, 160, 640},
    tick label style={font=\scriptsize},
    x tick label style={yshift=0.5ex},
    legend style={
        at={(-0.4,1.09)},anchor=south west,column sep=0pt,
        legend image post style={xscale=0.6},
        row sep = -0.4pt,
        draw=black,fill=white,
        inner ysep=0.1pt,
        /tikz/every even column/.append style={column sep=2pt},
        font=\footnotesize
    },
    legend cell align={left},
    legend columns=4,
    ylabel={recall},
    ymajorgrids,
    every axis plot/.append style={thick}
    % legend image code/.code={%
    % \draw[#1, draw=none] (0cm,-0.1cm) rectangle (0.6cm,0.1cm);}
]

% \addplot[line width=1pt, GreenColor,mark=*] coordinates
% {(1, 51.2) (2, 47.7) (3, 50.5) (4, 62.2) (5, 63.0)(6, 75.2)};
% Read
\addplot[line width=1pt, HeuristicColor, densely dashed]
table[x=x,y=y] {
x y
0 0.5812 
1 0.7563 
2 0.8474  
3 0.9177 
4 0.9641  
5 0.9885 
6 0.9955  
};
\addlegendentry{\diskann (\gist)}
\addplot[line width=1pt, DiskAnnColor, densely dotted]
table[x=x,y=y] {
x y
0 0.8505
1 0.9447 
2 0.9733
3 0.9882
4 0.9961  
5 0.9989
};
\addlegendentry{\diskann (\deep)}
% ## 4times

\end{axis}
\end{tikzpicture}
\vspace{-6.5mm}
\caption{search\_len vs. recall}
\label{fig:exp_diskann_deep_recall}
\end{subfigure}
\begin{subfigure}[b]{0.32\linewidth}
\begin{tikzpicture}

\begin{axis}[
    xtick=data,
    width=32mm,
    height=28mm,
    ymin=0,
    ymax=3,
    axis y line*=none,
    axis x line*=none,
    ytick={0, 1, 2, 3},
    yticklabels={0, 1, 2, 3},
    xlabel=Recall,
    xlabel style={yshift = 1.5ex},
    label style={font=\scriptsize},
    ylabel style={yshift=-1ex,xshift=-0.5ex, font=\scriptsize},
    xmin = 0.6,
    xmax = 1,
    xtick = {0.6, 0.7, 0.8, 0.9, 1.0},
    xticklabels = {0.6, 0.7, 0.8, 0.9, 1.0},
    tick label style={font=\scriptsize},
    x tick label style={yshift=0.5ex},
    legend style={
        at={(-0.6,1.09)},anchor=south west,column sep=0pt,
        legend image post style={xscale=0.6},
        row sep = -0.4pt,
        draw=black,fill=white,
        inner ysep=0.1pt,
        /tikz/every even column/.append style={column sep=2pt},
        font=\footnotesize
    },
    legend cell align={left},
    legend columns=4,
    ylabel={Roundtrips},
    ymajorgrids,
    every axis plot/.append style={thick}
    % legend image code/.code={%
    % \draw[#1, draw=none] (0cm,-0.1cm) rectangle (0.6cm,0.1cm);}
]

% \addplot[line width=1pt, GreenColor,mark=*] coordinates
% {(1, 51.2) (2, 47.7) (3, 50.5) (4, 62.2) (5, 63.0)(6, 75.2)};
% Read
\addplot[line width=1pt, HeuristicColor, densely dashed]
table[x=x,y=y] {
x y
0.5812  0.157
0.7563  0.275
0.8474  0.346
0.9177  0.477
0.9641  0.777
0.9885  1.411
0.9955  2.703
};
\addplot[line width=1pt, DiskAnnColor, densely dotted]
table[x=x,y=y] {
x y
0.8505 0.23
0.9447 0.37
0.9733 0.43
0.9882 0.658
0.9961 0.88
0.9989 1.52
};
% ## 4times

\end{axis}
\end{tikzpicture}
\vspace{-2.5mm}
\caption{Data Read per Query}
\label{fig:exp_diskann_deep_dataread}
\end{subfigure}
\begin{subfigure}[b]{0.32\linewidth}
\begin{tikzpicture}

\begin{axis}[
    xtick=data,
    width=32mm,
    height=28mm,
    ymin=0,
    ymax=50,
    axis y line*=none,
    axis x line*=none,
    ytick={0, 10, 20, 30, 40, 50},
    yticklabels={0, 10, 20, 30, 40, 50},
    xlabel=Recall,
    xlabel style={yshift = 1.5ex},
    label style={font=\scriptsize},
    ylabel style={yshift=-1ex,xshift=-0.5ex, font=\scriptsize},
    xmin = 0.6,
    xmax = 1,
    xtick = {0.6, 0.7, 0.8, 0.9, 1.0},
    xticklabels = {0.6, 0.7, 0.8, 0.9, 1.0},
    tick label style={font=\scriptsize},
    x tick label style={yshift=0.5ex},
    legend style={
        at={(-0.6,1.09)},anchor=south west,column sep=0pt,
        legend image post style={xscale=0.6},
        row sep = -0.4pt,
        draw=black,fill=white,
        inner ysep=0.1pt,
        /tikz/every even column/.append style={column sep=2pt},
        font=\footnotesize
    },
    legend cell align={left},
    legend columns=4,
    ylabel={Roundtrips},
    ymajorgrids,
    every axis plot/.append style={thick}
    % legend image code/.code={%
    % \draw[#1, draw=none] (0cm,-0.1cm) rectangle (0.6cm,0.1cm);}
]

% \addplot[line width=1pt, GreenColor,mark=*] coordinates
% {(1, 51.2) (2, 47.7) (3, 50.5) (4, 62.2) (5, 63.0)(6, 75.2)};
% Read
\addplot[line width=1pt, HeuristicColor, densely dashed]
table[x=x,y=y] {
x y
0.5812  7
0.7563  7
0.8474  8
0.9177  10
0.9641  14
0.9885  24
0.9955  43
};
\addplot[line width=1pt, DiskAnnColor, densely dotted]
table[x=x,y=y] {
x y
0.8505 8
0.9447 8
0.9733 9
0.9882 11
0.9961 15
0.9989 25
};
% ## 4times

\end{axis}
\end{tikzpicture}
\vspace{-2.5mm}
\caption{Roundtrips per Query}
\label{fig:exp_diskann_deep_roundtrips}
\end{subfigure}
\vspace{-1mm}
\caption{\diskann on \deep: lower-dimension dataset reduces \texttt{search\_len} values for certain recalls and consequently roundtrips, which \diskann's QPS benefits (one-fold) from.
}
\vspace{-3mm}
\label{fig:exp_diskann_deep}
\end{figure}
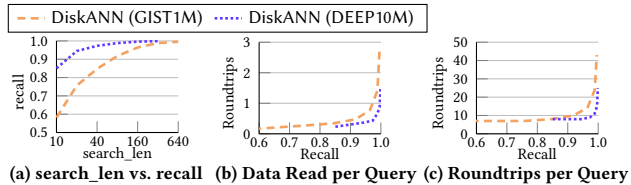
\paragraph{\deep: Low-Dimension Datasets Benefit \spann}
\deep features 96D vectors instead of \gist's 960D vectors (\cref{tbl:workload}), and also a higher recall/concurrency cutoff for \diskann to outperform \spann at 0.99 recall with 16 concurrent queries (\cref{fig:exp_spann_vs_diskann_deep_concurrency16}). These two factors are related: \spann benefits two-fold from querying on a low-dimension dataset (\Cref{fig:exp_spann_deep}): the same recall can be reached with a lower \texttt{nprobe} value, with only \texttt{nprobe=512} required to reach 0.995 recall on \deep versus \gist's 2048 (\cref{fig:exp_spann_deep_recall}, also empirically verified by Milvus~\cite{milvusblog}), and each posting list, containing the same number of significantly lower-dimension vectors,\footnote{Recall that \spann uses a \textit{percentage} of points as centroids.} are significantly smaller---15.0KB on \deep versus 166KB on \gist. These two factors combined significantly reduce the data read per query at the same recall by up to 28.4$\times$ (\cref{fig:exp_spann_deep_dataread}) and consequently improve QPS by up to 7.61$\times$ at 0.995 recall (concurrency=4, \cref{fig:exp_spann_vs_diskann_deep_concurrency4}). Notably, despite the reduction in data read, \spann's QPS on \deep is still bottlenecked by remote storage's GET QPS at high recalls and concurrencies (\cref{fig:exp_spann_deep_getqps}): despite the I/O bandwidth only being at 283 MB/s at concurrency=64 and 0.995 recall (\cref{fig:exp_spann_deep_bw}), the QPS improves by only 1.12$\times$ from concurrency=16 to 64 (\cref{fig:exp_spann_vs_diskann_deep}).

In contrast, \diskann benefits only one-fold from querying on \deep; while the same recall can similarly be reached with a lower \texttt{search\_len} value---160 on \deep vs. 640 on \gist for 0.995 recall (\cref{fig:exp_diskann_deep_recall}), which reduces data read per query by 3.07$\times$ (\cref{fig:exp_diskann_deep_dataread}) and  roundtrips by 2.88$\times$ (43 to 15 ,\cref{fig:exp_diskann_deep_roundtrips}) at this recall, none of these metrics further benefit from the reduced vector dimensions as \diskann reads data in fixed 4KB blocks~\cite{diskanngithub}. Hence, the QPS of \diskann only improves by up to 2.86$\times$ at 0.995 recall from \gist to \deep (concurrency=1, \cref{fig:exp_spann_vs_diskann_deep_concurrency1}).

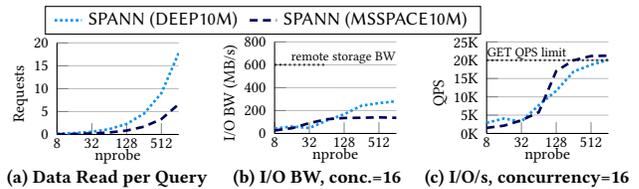
\begin{figure}[t]\captionsetup[subfigure]{font=footnotesize}
\pgfplotsset{scaled y ticks=false}
\centering
\begin{subfigure}[b]{0.32\linewidth}
\begin{tikzpicture}

\begin{axis}[
    xtick=data,
    width=32mm,
    height=28mm,
    ymin=0,
    ymax=20,
    axis y line*=none,
    axis x line*=none,
    ytick={0, 5, 10, 15, 20},
    yticklabels={0, 5, 10, 15, 20},
    xlabel=nprobe,
    xlabel style={yshift = 1.5ex},
    label style={font=\scriptsize},
    ylabel style={yshift=-1ex,xshift=-0.5ex, font=\scriptsize},
    xmin = 0,
    xmax = 7,
    xtick={0, 2, 4, 6},
    xticklabels={8, 32, 128, 512},
    tick label style={font=\scriptsize},
    x tick label style={yshift=0.5ex},
    legend style={
        at={(-0.1,1.09)},anchor=south west,column sep=0pt,
        legend image post style={xscale=0.6},
        row sep = -0.4pt,
        draw=black,fill=white,
        inner ysep=0.1pt,
        /tikz/every even column/.append style={column sep=2pt},
        font=\footnotesize
    },
    legend cell align={left},
    legend columns=4,
    ylabel={Requests},
    ymajorgrids,
    every axis plot/.append style={thick}
    % legend image code/.code={%
    % \draw[#1, draw=none] (0cm,-0.1cm) rectangle (0.6cm,0.1cm);}
]

% \addplot[line width=1pt, GreenColor,mark=*] coordinates
% {(1, 51.2) (2, 47.7) (3, 50.5) (4, 62.2) (5, 63.0)(6, 75.2)};
% Read
\addplot[line width=1pt, AcornOneColor, densely dotted]
table[x=x,y=y] {
x y
0 0.149
1 0.296 
2 0.586 
3 1.161 
4 2.298
5 4.549
6 9.006
7 17.829
};
\addlegendentry{\spann (\deep)}
\addplot[line width=1pt, AcornGammaColor, densely dashed]
table[x=x,y=y] {
x y
0 0.055
1 0.110 
2 0.218
3 0.432 
4 0.855
5 1.694
6 3.353
7 6.619
};
\addlegendentry{\spann (\msspace)}
% ## 4times

\end{axis}
\end{tikzpicture}
\vspace{-6.5mm}
\caption{Data Read per Query}
\label{fig:exp_spann_msspace_recall}
\end{subfigure}
\begin{subfigure}[b]{0.32\linewidth}
\begin{tikzpicture}

\begin{axis}[
    xtick=data,
    width=32mm,
    height=28mm,
    ymin=0,
    ymax=800,
    axis y line*=none,
    axis x line*=none,
    ytick={0, 200, 400, 600, 800},
    yticklabels={0, 200, 400, 600, 800},
    xlabel=nprobe,
    xlabel style={yshift = 1.5ex},
    label style={font=\scriptsize},
    ylabel style={yshift=-1ex,xshift=-0.5ex, font=\scriptsize},
    xmin = 0,
    xmax = 7,
    xtick={0, 2, 4, 6},
    xticklabels={8, 32, 128, 512},
    tick label style={font=\scriptsize},
    x tick label style={yshift=0.5ex},
    legend style={
        at={(-0.4,1.09)},anchor=south west,column sep=0pt,
        legend image post style={xscale=0.6},
        row sep = -0.4pt,
        draw=black,fill=white,
        inner ysep=0.1pt,
        /tikz/every even column/.append style={column sep=2pt},
        font=\footnotesize
    },
    legend cell align={left},
    legend columns=4,
    ylabel={I/O BW (MB/s)},
    ymajorgrids,
    every axis plot/.append style={thick}
    % legend image code/.code={%
    % \draw[#1, draw=none] (0cm,-0.1cm) rectangle (0.6cm,0.1cm);}
]

% \addplot[line width=1pt, GreenColor,mark=*] coordinates
% {(1, 51.2) (2, 47.7) (3, 50.5) (4, 62.2) (5, 63.0)(6, 75.2)};
% Read
\addplot[line width=1pt, AcornOneColor, densely dotted]
table[x=x,y=y] {
x y
0 43.622
1 60.586
2 47.427
3 112.678
4 167.712
5 242.683
6 265.191
7 281.224
};
\addplot[line width=1pt, AcornGammaColor, densely dashed]
table[x=x,y=y] {
x y
0 26.412
1 45.980
2 90.613
3 124.135
4 135.163
5 137.332
6 140.435
7 136.826
};
% ## 4times
\draw[black, thick, opacity=0.7, densely dotted] (axis cs: 0, 600) -- (axis cs: 3, 600);
\node[anchor=south east, align=right, inner sep = 0.1pt] at (axis cs: 7, 600) {\tiny remote storage BW};
\end{axis}
\end{tikzpicture}
\vspace{-2.5mm}
\caption{I/O BW, conc.=16}
\label{fig:exp_spann_msspace_bw}
\end{subfigure}
\begin{subfigure}[b]{0.32\linewidth}
\begin{tikzpicture}

\begin{axis}[
    xtick=data,
    width=32mm,
    height=28mm,
    ymin=0,
    ymax=25,
    axis y line*=none,
    axis x line*=none,
    ytick={0, 5, 10, 15, 20, 25},
    yticklabels={0K, 5K, 10K, 15K, 20K, 25K},
    xlabel=nprobe,
    xlabel style={yshift = 1.5ex},
    label style={font=\scriptsize},
    ylabel style={yshift=-1ex,xshift=-0.5ex, font=\scriptsize},
    xmin = 0,
    xmax = 7,
    xtick={0, 2, 4, 6},
    xticklabels={8, 32, 128, 512},
    tick label style={font=\scriptsize},
    x tick label style={yshift=0.5ex},
    legend style={
        at={(-0.4,1.09)},anchor=south west,column sep=0pt,
        legend image post style={xscale=0.6},
        row sep = -0.4pt,
        draw=black,fill=white,
        inner ysep=0.1pt,
        /tikz/every even column/.append style={column sep=2pt},
        font=\footnotesize
    },
    legend cell align={left},
    legend columns=4,
    ylabel={QPS},
    ymajorgrids,
    every axis plot/.append style={thick}
    % legend image code/.code={%
    % \draw[#1, draw=none] (0cm,-0.1cm) rectangle (0.6cm,0.1cm);}
]

% \addplot[line width=1pt, GreenColor,mark=*] coordinates
% {(1, 51.2) (2, 47.7) (3, 50.5) (4, 62.2) (5, 63.0)(6, 75.2)};
% Read
\addplot[line width=1pt, AcornOneColor, densely dotted]
table[x=x,y=y] {
x y
0 2.910
1 4.080
2 3.226
3 7.736
4 11.630
5 16.890
6 18.732
7 20.099
};
\addplot[line width=1pt, AcornGammaColor, densely dashed]
table[x=x,y=y] {
x y
0 1.528
1 2.126
2 3.592
3 5.809
4 17.023
5 19.972
6 21.136
7 21.228
};
% ## 4times
\draw[black, thick, opacity=0.7, densely dotted] (axis cs: 0, 20) -- (axis cs: 7, 20);
\node[anchor=south west, align=right, inner sep = 0.1pt] at (axis cs: 0, 20) {\tiny GET QPS limit};
\end{axis}
\end{tikzpicture}
\vspace{-2.5mm}
\caption{I/O/s, concurrency=16}
\label{fig:exp_spann_msspace_ios}
\end{subfigure}
\vspace{-1mm}
\caption{\spann on \msspace: quantized datatypes (e.g., INT8) reduces posting list size and data read per query, improving QPS when the GET QPS limit is yet to be reached.
}
\vspace{-3mm}
\label{fig:exp_spann_msspace}
\end{figure}
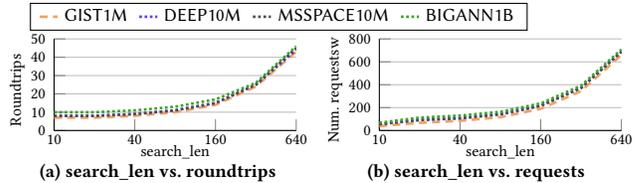
\begin{figure}[t]\captionsetup[subfigure]{font=footnotesize}
\pgfplotsset{scaled y ticks=false}
\centering
\begin{subfigure}[b]{0.49\linewidth}
\begin{tikzpicture}

\begin{axis}[
    xtick=data,
    width=48mm,
    height=28mm,
    ymin=0,
    ymax=50,
    axis y line*=none,
    axis x line*=none,
    ytick={0, 10, 20, 30, 40, 50},
    yticklabels={0, 10, 20, 30, 40, 50},
    xlabel=search\_len,
    xlabel style={yshift = 1.5ex},
    label style={font=\scriptsize},
    ylabel style={yshift=-1ex,xshift=-0.5ex, font=\scriptsize},
    xmin = 0,
    xmax = 6,
    xtick={0, 2, 4, 6},
    xticklabels={10, 40, 160, 640},
    tick label style={font=\scriptsize},
    x tick label style={yshift=0.5ex},
    legend style={
        at={(-0.1,1.09)},anchor=south west,column sep=0pt,
        legend image post style={xscale=0.4},
        row sep = -0.4pt,
        draw=black,fill=white,
        inner ysep=0.1pt,
        /tikz/every even column/.append style={column sep=2pt},
        font=\footnotesize
    },
    legend cell align={left},
    legend columns=4,
    ylabel={Roundtrips},
    ymajorgrids,
    every axis plot/.append style={thick}
    % legend image code/.code={%
    % \draw[#1, draw=none] (0cm,-0.1cm) rectangle (0.6cm,0.1cm);}
]

% \addplot[line width=1pt, GreenColor,mark=*] coordinates
% {(1, 51.2) (2, 47.7) (3, 50.5) (4, 62.2) (5, 63.0)(6, 75.2)};
% Read
\addplot[line width=1pt, HeuristicColor, densely dashed]
table[x=x,y=y] {
x y
0 7
1  7
2  8
3  10
4  14
5  24
6  43
};
\addlegendentry{\gist}
\addplot[line width=1pt, DiskAnnColor, densely dotted]
table[x=x,y=y] {
x y
0  8
1 8
2  9
3  11
4  15
5 25
6 44
};
\addlegendentry{\deep}
\addplot[line width=1pt, PreFilterColor, densely dotted]
table[x=x,y=y] {
x y
0  8
1 8
2  9
3  11
4  15
5 25
6 45
};
\addlegendentry{\msspace}
\addplot[line width=1pt, CapsColor, densely dotted]
table[x=x,y=y] {
x y
0 10
1 10
2 11
3 13
4 17
5 26
6 46
};
\addlegendentry{\bigann}
% ## 4times

\end{axis}
\end{tikzpicture}
\vspace{-6.5mm}
\caption{search\_len vs. roundtrips}
\label{fig:exp_diskann_msspace_searchlen}
\end{subfigure}
\begin{subfigure}[b]{0.49\linewidth}
\begin{tikzpicture}

\begin{axis}[
    xtick=data,
    width=48mm,
    height=28mm,
    ymin=0,
    ymax=800,
    axis y line*=none,
    axis x line*=none,
    ytick={0, 200, 400, 600, 800},
    yticklabels={0, 200, 400, 600, 800},
    xlabel=search\_len,
    xlabel style={yshift = 1.5ex},
    label style={font=\scriptsize},
    ylabel style={yshift=-1ex,xshift=-0.5ex, font=\scriptsize},
    xmin = 0,
    xmax = 6,
    xtick={0, 2, 4, 6},
    xticklabels={10, 40, 160, 640},
    tick label style={font=\scriptsize},
    x tick label style={yshift=0.5ex},
    legend style={
        at={(-0.1,1.09)},anchor=south west,column sep=0pt,
        legend image post style={xscale=0.4},
        row sep = -0.4pt,
        draw=black,fill=white,
        inner ysep=0.1pt,
        /tikz/every even column/.append style={column sep=2pt},
        font=\footnotesize
    },
    legend cell align={left},
    legend columns=4,
    ylabel={Num. requestsw},
    ymajorgrids,
    every axis plot/.append style={thick}
    % legend image code/.code={%
    % \draw[#1, draw=none] (0cm,-0.1cm) rectangle (0.6cm,0.1cm);}
]

% \addplot[line width=1pt, GreenColor,mark=*] coordinates
% {(1, 51.2) (2, 47.7) (3, 50.5) (4, 62.2) (5, 63.0)(6, 75.2)};
% Read
\addplot[line width=1pt, HeuristicColor, densely dashed]
table[x=x,y=y] {
x y
0  38.45
1  67.19
2  84.58
3  116.56
4  189.85
5  344.5
6  660.16
};
\addplot[line width=1pt, DiskAnnColor, densely dotted]
table[x=x,y=y] {
x y
0  55.75
1  91.49
2  106.39
3  140.73
4  216.34
5  372.44
6  688.9
};
\addplot[line width=1pt, PreFilterColor, densely dotted]
table[x=x,y=y] {
x y
0  55.39
1 92.21
2  108.16
3  142.4
4  216.78
5 371.63
6 687.03
};
\addplot[line width=1pt, CapsColor, densely dotted]
table[x=x,y=y] {
x y
0 67.55
1 112.65
2 130.1
3 163.07
4 237.03
5 392.12
6 707.87
};
% ## 4times

\end{axis}
\end{tikzpicture}
\vspace{-6.5mm}
\caption{search\_len vs. requests}
\label{fig:exp_diskann_msspace_num_requests}
\end{subfigure}
\vspace{-1mm}
\caption{\diskann's average roundtrips and number of requests per query versus dataset size: both of these metrics scale logarithmically versus dataset size; \diskann'ss query latency is significantly correlated with the former.
}
\vspace{-3mm}
\label{fig:exp_diskann_msspace}
\end{figure}
\paragraph{\msspace: Quantized Datasets Benefit \spann}
\msspace features 100D vectors of quantized INT8 datatype, which when compared to the similar-dimension \deep using FLOAT32, has $\sim4\times$ smaller-sized vectors (\cref{tbl:workload}). \msspace features an even higher recall/concurrency cutoff for \diskann to outperform \spann at 0.96 recall and 64 concurrent queries (\cref{fig:exp_spann_vs_diskann_msspace_concurrency64}). This is because \spann benefits from the smaller-size INT8 datatype on \msspace, reading a uniform 2.69$\times$ less data per query versus \deep at every \texttt{nprobe} value (\cref{fig:exp_spann_msspace_recall}) allowing for better utilization of I/O bandwidth (\cref{fig:exp_spann_msspace_bw}). This significantly improves QPS at lower \texttt{nprobe} values (up to 2.13$\times$ at \texttt{nprobe}=8, \cref{fig:exp_spann_vs_diskann_msspace_concurrency16}), with the improvement diminishing as the GET QPS limit is approached (\cref{fig:exp_spann_msspace_ios}), down to only up to 1.05$\times$ at \texttt{nprobe=1028} (\cref{fig:exp_spann_vs_diskann_msspace_concurrency64}).\footnote{Different datasets achieve different recalls with different \texttt{nprobe} values. We use \texttt{nprobe} vs. QPS to isolate the effect of datatype.} Due to fixed block read sizes, \diskann does not benefit from quantized datatypes QPS-wise given fixed \texttt{search\_len} values.
% This is because 
% \Cref{fig:exp_spann_deep} and \cref{fig:exp_diskann_deep} report results of \spann and \diskann on \deep built following \cref{tbl:parameters}. \deep features 96D vectors instead of \gist's 960D vectors.

\paragraph{\bigann: Large Datasets Benefit \spann}
\bigann features 1 billion 128D INT8 vectors, and has the highest recall/concurrency cutoff for \diskann to outperform \spann, at 0.99 recall with 64 concurrent queries (\cref{fig:exp_spann_vs_diskann_bigann_concurrency64}). This is because while both indexes contains components that exhibit logarithimic scaling with dataset size---\spann's BKT tree traversal time (\cref{sec:cluster_index_cost}), and \diskann's number of roundtrips~\cite{zhi2025towards} (\cref{fig:exp_diskann_msspace_searchlen}) and requests per query~\cite{indyk2023worst} (\cref{fig:exp_diskann_msspace_num_requests}), \spann's BKT tree search is performed entirely in memory and comprises a small portion of the overall search cost (\cref{tbl:spann_overhead_metrics}), while \diskann's metrics, especially number of roundtrips, is core to the query latency (\cref{fig:cache_gains_diskann}). 

\begin{tcolorbox}[colback=gray!10,colframe=gray!40,boxrule=0.5pt,arc=2pt,left=5pt,right=5pt,top=5pt,bottom=5pt]
\paragraph{A1: What index for what scenario?} In general, \spann features lower query latency versus \diskann, at the cost of higher data read per query, and more I/Os per query at high recalls. Consequently, under typical cloud setups, \diskann outperforms \spann at high query recalls and/or concurrencies and vice versa. Dataset characteristics will affect balance: low vector dataset dimensionality, small (e.g., quantized) datatypes, and larger dataset size shift the balance in favor of \spann.
\end{tcolorbox}
\subsection{How to Design Indexes?}
\label{sec:exp_parameterization}
This section studies how \spann and \diskann can be adjusted for cloud-native vector search. We explore how various parameterizations, even those outside the recommended range for on-disk querying, may achieve higher on-cloud query performance according to our cost models (\cref{sec:cloud_index}) on the \gist dataset.

% \subsubsection{Tuning \spann}
\paragraph{\spann: More Centroids Benefit I/O-Congested Setups}
\Cref{fig:exp_spann_more_centroids_qps} reports the QPS ratio of an alternative \spann index built with a higher centroid\%=32 versus the default \spann index built with centroid\%=16. As hypothesized in \cref{sec:cloud_index_improve}, the former achieves QPS gains versus the latter on high recall and/or concurrency scenarios (up to 3.14$\times$) as the centroid\%=32 count index contains more posting lists each of significantly smaller size (by 32.1\%, \cref{tbl:spann_index_size}),\footnote{\spann contains optimizations for pruning redundant centroids~\cite{chen2021spann}, hence the actual number of centroids is less than 32\%~.} which significantly reduces the amount of data read per query (1.47$\times$ at 0.995 recall, \cref{fig:exp_spann_more_centroids_metrics_dataread}). The network metrics at 0.995 recall affirm this: while bandwidth is still saturated (\cref{fig:exp_spann_more_centroids_metrics_bw}), the mean I/O latency has decreased by 1.39$\times$ at 0.995 recall with concurrency=64 (\cref{fig:exp_spann_more_centroids_metrics_meanio}).\footnote{For each \texttt{nprobe} value, the centroid\%=32 achieves similar recalls versus the centroid\%=16 index ($\pm$1\%). We treat them as identical for brevity.}

However, the centroid\%=32 index underperforms versus the centroid\%=16 index on low concurrency and/or recall scenarios where the I/O congestion is low (\cref{fig:exp_meanio_gist}). As previously shown in \cref{tbl:spann_index_tuning}, despite saving I/O costs, the centroid\%=32 index incurs higher distance computations and BKT Tree traversal costs, losing performance when BKT tree search overhead ($O(logn)$) is more significant than that of I/O cost ($O(n)$), matching conventional wisdom of using centroid counts of under 20\% for on-disk querying~\cite{chen2021spann}. 
\begin{figure}[t]\captionsetup[subfigure]{font=footnotesize}
\pgfplotsset{scaled y ticks=false}
\centering

\begin{subfigure}[b]{\linewidth}
\begin{tikzpicture}
\begin{axis}[
    width=7cm, height=5cm,
    matrix plot,
    colormap/viridis,
    xlabel={Number of Concurrent Queries},
    ylabel={nprobe / Recall},
    xtick={1,2,3,4},
    xticklabels={1, 4, 16, 64},
    ytick={1,2,3,4,5,6,7,8,9},
    yticklabels={
        8 (65.25\%), 
        16 (72.45\%), 
        32 (80.33\%), 
        64 (87.67\%), 
        128 (93.34\%), 
        256 (96.90\%), 
        512 (98.49\%), 
        1024 (99.25\%), 
        2048 (99.63\%)
    },
    label style={font=\scriptsize},
    tick label style={font=\scriptsize},
    colorbar,
    point meta min=0.3,
    point meta max=2.0,
    nodes near coords,
    nodes near coords align={center},
    every node near coord/.append style={font=\scriptsize, color=black, opacity=0.9, fill=white, text opacity=1, rounded corners=2pt, inner sep=0.8pt},
    y dir=reverse,
    ]
\addplot [matrix plot*, point meta=explicit] table [meta=ratio] {
x y ratio
1 1 0.695
2 1 0.344
3 1 0.986
4 1 1.148

1 2 0.626
2 2 0.792
3 2 0.636
4 2 1.014

1 3 0.660
2 3 1.282
3 3 1.619
4 3 3.141

1 4 0.705
2 4 1.223
3 4 1.950
4 4 1.583

1 5 1.213
2 5 1.251
3 5 1.911
4 5 1.405

1 6 1.428
2 6 1.534
3 6 1.503
4 6 1.491

1 7 1.657
2 7 1.558
3 7 2.490
4 7 1.401

1 8 1.121
2 8 1.526
3 8 1.784
4 8 1.337

1 9 1.299
2 9 1.532
3 9 1.420
4 9 1.443
};
\end{axis}
\end{tikzpicture}
\vspace{-4.5mm}
\caption{Ratio of (centroid\%=32 QPS/centroid\%=16 QPS of \spann on \gist}
\end{subfigure}
\vspace{-5mm}
\caption{QPS ratio of the alternative \spann index with fine-grained posting lists (centroid\%=32) over the default centroid\%=16 index. The alternative achieves significant QPS increases on high recall and/or concurrency scenarios.
}
\vspace{-3mm}
\label{fig:exp_spann_more_centroids_qps}
\end{figure}
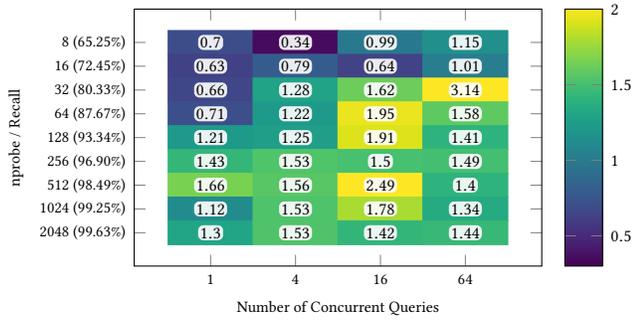
\begin{table}[t]

\caption{Size metrics of \spann configurations on \gist.}
\footnotesize
\vspace{-1mm}
\addtolength{\tabcolsep}{-1.5pt} 
\midsepremove
\begin{tabular}{|l|c|c|c|}
\toprule
 % & \multicolumn{16}{c|}{Indexing Method}\\
 %  \midrule
Configuration & Index size (GB) & No. lists  & Avg. list size (KB) \\
 \hline
\spann (centroid\%=16,replica=8) & 13.0 & 159K & 166 \\
\hline
\spann (centroid\%=16,replica=4) & 10.5 & 159K & 138 \\
\hline
\spann (centroid\%=16,replica=2) & \textbf{7.5} & 159K & \textbf{99} \\
\hline
\spann (centroid\%=32,replica=8) & 14.0 & 271K & 119 \\
% \hline
% \# Exact searches@$efc=10$ & \textbf{3907} & 3821 & \textbf{2697} & 2099  \\
\bottomrule
\end{tabular}
\vspace{-3mm}
\midsepdefault
\addtolength{\tabcolsep}{1.5pt}
\label{tbl:spann_index_size}
\end{table}

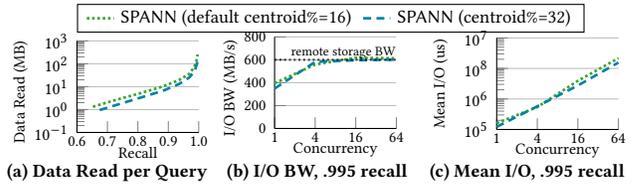
\begin{figure}[t]\captionsetup[subfigure]{font=footnotesize}
\pgfplotsset{scaled y ticks=false}
\centering
\begin{subfigure}[b]{0.32\linewidth}
\begin{tikzpicture}

\begin{axis}[
    xtick=data,
    width=32mm,
    height=28mm,
    ymin=0.1,
    ymax=1000,
    log origin = infty,
    ymode = log,
    axis y line*=none,
    axis x line*=none,
    ytick={0.1, 1, 10, 100, 1000},
    yticklabels={$10^{-1}$, $10^0$, $10^1$, $10^2$,$10^3$},
    xlabel=Recall,
    xlabel style={yshift = 1.5ex},
    label style={font=\scriptsize},
    ylabel style={yshift=-1ex,xshift=-0.5ex, font=\scriptsize},
    xmin = 0.6,
    xmax = 1,
    xtick = {0.6, 0.7, 0.8, 0.9, 1.0},
    xticklabels = {0.6, 0.7, 0.8, 0.9, 1.0},
    tick label style={font=\scriptsize},
    x tick label style={yshift=0.5ex},
    legend style={
        at={(0,1.09)},anchor=south west,column sep=0pt,
        legend image post style={xscale=0.6},
        row sep = -0.4pt,
        draw=black,fill=white,
        inner ysep=0.1pt,
        /tikz/every even column/.append style={column sep=2pt},
        font=\footnotesize
    },
    legend cell align={left},
    legend columns=4,
    ylabel={Data Read (MB)},
    ymajorgrids,
    every axis plot/.append style={thick}
    % legend image code/.code={%
    % \draw[#1, draw=none] (0cm,-0.1cm) rectangle (0.6cm,0.1cm);}
]

% \addplot[line width=1pt, GreenColor,mark=*] coordinates
% {(1, 51.2) (2, 47.7) (3, 50.5) (4, 62.2) (5, 63.0)(6, 75.2)};
% Read
\addplot[line width=1pt, GreenColor, densely dotted]
table[x=x,y=y] {
x y
0.6525  1.331
0.7245  2.578
0.8033  4.979
0.8767  9.708
0.9334  18.827
0.9690  36.447
0.9849  70.349
0.9925  134.396
0.9963  256.214
};
\addlegendentry{\spann (default centroid\%=16)}
\addplot[line width=1pt, BlueColor, densely dashed]
table[x=x,y=y] {
x y
0.6746 0.951
0.7446 1.835
0.8157 3.54
0.8851 6.842
0.936 13.203
0.9692 25.335
0.9866 48.429
0.9935 92.344
0.9968 174.165
};
\addlegendentry{\spann (centroid\%=32)}
% ## 4times

\end{axis}
\end{tikzpicture}
\vspace{-6.5mm}
\caption{Data Read per Query}
\label{fig:exp_spann_more_centroids_metrics_dataread}
\end{subfigure}
\begin{subfigure}[b]{0.32\linewidth}
\begin{tikzpicture}

\begin{axis}[
    xtick=data,
    width=32mm,
    height=28mm,
    ymin=0,
    ymax=800,
    axis y line*=none,
    axis x line*=none,
    ytick={0, 200, 400, 600, 800},
    yticklabels={0, 200, 400, 600, 800},
    xlabel=Concurrency,
    xlabel style={yshift = 1.5ex},
    label style={font=\scriptsize},
    ylabel style={yshift=-1ex,xshift=-0.5ex, font=\scriptsize},
    xmin = 0,
    xmax = 3,
    xtick = {0, 1, 2, 3},
    xticklabels = {1, 4, 16, 64},
    tick label style={font=\scriptsize},
    x tick label style={yshift=0.5ex},
    legend style={
        at={(-0.4,1.09)},anchor=south west,column sep=0pt,
        legend image post style={xscale=0.6},
        row sep = -0.4pt,
        draw=black,fill=white,
        inner ysep=0.1pt,
        /tikz/every even column/.append style={column sep=2pt},
        font=\footnotesize
    },
    legend cell align={left},
    legend columns=4,
    ylabel={I/O BW (MB/s)},
    ymajorgrids,
    every axis plot/.append style={thick}
    % legend image code/.code={%
    % \draw[#1, draw=none] (0cm,-0.1cm) rectangle (0.6cm,0.1cm);}
]

% \addplot[line width=1pt, GreenColor,mark=*] coordinates
% {(1, 51.2) (2, 47.7) (3, 50.5) (4, 62.2) (5, 63.0)(6, 75.2)};
% Read
\addplot[line width=1pt, GreenColor, densely dotted]
table[x=x,y=y] {
x y
0 395.4634
1 558.7701
2 618.1325
3 613.0984
};
\addplot[line width=1pt, BlueColor, densely dashed]
table[x=x,y=y] {
x y
0 349.0192
1 581.8225
2 596.5788
3 601.4001
};
\draw[black, thick, opacity=0.7, densely dotted] (axis cs: 0, 600) -- (axis cs: 3, 600);
\node[anchor=south east, align=right, inner sep = 0.1pt] at (axis cs: 3, 600) {\tiny remote storage BW};
% ## 4times

\end{axis}
\end{tikzpicture}
\vspace{-2.5mm}
\caption{I/O BW, .995 recall}
\label{fig:exp_spann_more_centroids_metrics_bw}
\end{subfigure}
\begin{subfigure}[b]{0.32\linewidth}
\begin{tikzpicture}

\begin{axis}[
    xtick=data,
    width=32mm,
    height=28mm,
    ymin=100000,
    ymax=100000000,
    log origin = infty,
    ymode = log,
    axis y line*=none,
    axis x line*=none,
    ytick={100000, 1000000, 10000000, 100000000},
    yticklabels={$10^5$, $10^6$, $10^8$, $10^7$},
    xlabel=Concurrency,
    xlabel style={yshift = 1.5ex},
    xmin = 0,
    xmax = 3,
    xtick = {0, 1, 2, 3},
    xticklabels = {1, 4, 16, 64},
    tick label style={font=\scriptsize},
    label style={font=\scriptsize},
    x tick label style={yshift=0.5ex},
    legend style={
        at={(-0.4,1.09)},anchor=south west,column sep=0pt,
        legend image post style={xscale=0.6},
        row sep = -0.4pt,
        draw=black,fill=white,
        inner ysep=0.1pt,
        /tikz/every even column/.append style={column sep=2pt},
        font=\footnotesize
    },
    legend cell align={left},
    legend columns=4,
    ylabel={Mean I/O (us)},
    ymajorgrids,
    every axis plot/.append style={thick}
    % legend image code/.code={%
    % \draw[#1, draw=none] (0cm,-0.1cm) rectangle (0.6cm,0.1cm);}
]

% \addplot[line width=1pt, GreenColor,mark=*] coordinates
% {(1, 51.2) (2, 47.7) (3, 50.5) (4, 62.2) (5, 63.0)(6, 75.2)};
% Read
\addplot[line width=1pt, GreenColor, densely dotted]
table[x=x,y=y] {
x y
0 161423
1 542075
2 3933236
3 21641523
};
\addplot[line width=1pt, BlueColor, densely dashed]
table[x=x,y=y] {
x y
0 122061
1 573621
2 2765108
3 15606234
};

% ## 4times

\end{axis}
\end{tikzpicture}
\vspace{-6.5mm}
\caption{Mean I/O, .995 recall}
\label{fig:exp_spann_more_centroids_metrics_meanio}
\end{subfigure}
\vspace{-1mm}
\caption{\spann on \gist: Increasing the number of centroids and consequentially decreasing posting list size significantly reduces data read per query (left). While bandwidth is still saturated on high recall/concurrency scenarios (middle), congestion is significantly reduced (right).
}
\vspace{-3mm}
\label{fig:exp_spann_more_centroids_metrics}
\end{figure}
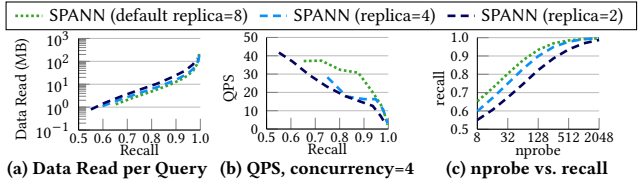
\begin{figure}[t]\captionsetup[subfigure]{font=footnotesize}
\pgfplotsset{scaled y ticks=false}
\centering
\begin{subfigure}[b]{0.32\linewidth}
\begin{tikzpicture}

\begin{axis}[
    xtick=data,
    width=32mm,
    height=28mm,
    ymin=0.1,
    ymax=1000,
    log origin = infty,
    ymode = log,
    axis y line*=none,
    axis x line*=none,
    ytick={0.1, 1, 10, 100, 1000},
    yticklabels={$10^{-1}$, $10^0$, $10^1$, $10^2$,$10^3$},
    xlabel=Recall,
    xlabel style={yshift = 1.5ex},
    label style={font=\scriptsize},
    ylabel style={yshift=-1ex,xshift=-0.5ex, font=\scriptsize},
    xmin = 0.5,
    xmax = 1,
    xtick = {0.5, 0.6, 0.7, 0.8, 0.9, 1.0},
    xticklabels = {0.5, 0.6, 0.7, 0.8, 0.9, 1.0},
    tick label style={font=\scriptsize},
    x tick label style={yshift=0.5ex},
    legend style={
        at={(-0.6,1.09)},anchor=south west,column sep=0pt,
        legend image post style={xscale=0.6},
        row sep = -0.4pt,
        draw=black,fill=white,
        inner ysep=0.1pt,
        /tikz/every even column/.append style={column sep=2pt},
        font=\footnotesize
    },
    legend cell align={left},
    legend columns=4,
    ylabel={Data Read (MB)},
    ymajorgrids,
    every axis plot/.append style={thick}
    % legend image code/.code={%
    % \draw[#1, draw=none] (0cm,-0.1cm) rectangle (0.6cm,0.1cm);}
]

% \addplot[line width=1pt, GreenColor,mark=*] coordinates
% {(1, 51.2) (2, 47.7) (3, 50.5) (4, 62.2) (5, 63.0)(6, 75.2)};
% Read
\addplot[line width=1pt, GreenColor, densely dotted]
table[x=x,y=y] {
x y
0.6525  1.331
0.7245  2.578
0.8033  4.979
0.8767  9.708
0.9334  18.827
0.9690  36.447
0.9849  70.349
0.9925  134.396
0.9963  256.214
};
\addlegendentry{\spann (default replica=8)}
\addplot[line width=1pt, AcornOneColor, densely dashed]
table[x=x,y=y] {
x y
0.5983 1.107
0.6708 2.139
0.7496 4.131
0.8296 7.963
0.8989 15.358
0.9488 29.38
0.978 56.098
0.9921 106.932
0.9962 201.982
};
\addlegendentry{\spann (replica=4)}
\addplot[line width=1pt, AcornGammaColor, densely dashed]
table[x=x,y=y] {
x y
0.5506 0.794
0.6059 1.514
0.6735 2.91 
0.7464 5.578
0.8231 10.668
0.8865 20.948
0.9359 38.397
0.9693 72.766
0.9873 135.46
};
\addlegendentry{\spann (replica=2)}
% ## 4times

\end{axis}
\end{tikzpicture}
\vspace{-6.5mm}
\caption{Data Read per Query}
\label{fig:exp_spann_lower_replica_metrics_dataread}
\end{subfigure}
\begin{subfigure}[b]{0.32\linewidth}
\begin{tikzpicture}

\begin{axis}[
    xtick=data,
    width=32mm,
    height=28mm,
    ymin=0,
    ymax=50,
    axis y line*=none,
    axis x line*=none,
    ytick={0, 10, 20, 30, 40, 50},
    yticklabels={0, 10, 20, 30, 40, 50},
    xlabel=Recall,
    xlabel style={yshift = 1.5ex},
    label style={font=\scriptsize},
    ylabel style={yshift=-1ex,xshift=-0.5ex, font=\scriptsize},
    xmin = 0.5,
    xmax = 1,
    xtick = {0.5, 0.6, 0.7, 0.8, 0.9, 1.0},
    xticklabels = {0.5, 0.6, 0.7, 0.8, 0.9, 1.0},
    tick label style={font=\scriptsize},
    x tick label style={yshift=0.5ex},
    legend style={
        at={(-0.4,1.09)},anchor=south west,column sep=0pt,
        legend image post style={xscale=0.6},
        row sep = -0.4pt,
        draw=black,fill=white,
        inner ysep=0.1pt,
        /tikz/every even column/.append style={column sep=2pt},
        font=\footnotesize
    },
    legend cell align={left},
    legend columns=4,
    ylabel={QPS},
    ymajorgrids,
    every axis plot/.append style={thick}
    % legend image code/.code={%
    % \draw[#1, draw=none] (0cm,-0.1cm) rectangle (0.6cm,0.1cm);}
]

% \addplot[line width=1pt, GreenColor,mark=*] coordinates
% {(1, 51.2) (2, 47.7) (3, 50.5) (4, 62.2) (5, 63.0)(6, 75.2)};
% Read
\addplot[line width=1pt, GreenColor, densely dotted]
table[x=x,y=y] {
x y
0.6525  37.1131
0.7245  37.3339
0.8033  32.2885
0.8767  30.7586
0.9334  20.5169
0.9690  13.4446
0.9849  8.034
0.9925  4.3294
0.9963  2.2869
};
\addplot[line width=1pt, AcornOneColor, densely dashed]
table[x=x,y=y] {
x y
% 0.5983 64.0783
% 0.6708 53.6224
0.7496 28.2705
0.8296 17.7507
0.8989 16.3505
0.9488 16.3012
0.978 10.8463
0.9921 5.2878
0.9962 2.9803
};
\addplot[line width=1pt, AcornGammaColor, densely dashed]
table[x=x,y=y] {
x y
0.5506 41.7428
0.6059 36.9154
0.6735 29.6889
0.7464 23.2851
0.8231 17.9751
0.8865 14.9556
0.9359 12.6315
0.9693 7.0076
0.9873 2.9124
};
% ## 4times

\end{axis}
\end{tikzpicture}
\vspace{-2.5mm}
\caption{QPS, concurrency=4}
\label{fig:exp_spann_lower_replica_metrics_qps}
\end{subfigure}
\begin{subfigure}[b]{0.32\linewidth}
\begin{tikzpicture}

\begin{axis}[
    xtick=data,
    width=32mm,
    height=28mm,
    ymin=0.5,
    ymax=1,
    axis y line*=none,
    axis x line*=none,
    ytick={0.5, 0.6, 0.7, 0.8, 0.9, 1.0},
    yticklabels={0.5, 0.6, 0.7, 0.8, 0.9, 1.0},
    xlabel=nprobe,
    xlabel style={yshift = 1.5ex},
    label style={font=\scriptsize},
    ylabel style={yshift=-1ex,xshift=-0.5ex, font=\scriptsize},
    xmin = 0,
    xmax = 8,
    xtick={0, 2, 4, 6, 8},
    xticklabels={8, 32, 128, 512, 2048},
    tick label style={font=\scriptsize},
    x tick label style={yshift=0.5ex},
    legend style={
        at={(-0.4,1.09)},anchor=south west,column sep=0pt,
        legend image post style={xscale=0.6},
        row sep = -0.4pt,
        draw=black,fill=white,
        inner ysep=0.1pt,
        /tikz/every even column/.append style={column sep=2pt},
        font=\footnotesize
    },
    legend cell align={left},
    legend columns=4,
    ylabel={recall},
    ymajorgrids,
    every axis plot/.append style={thick}
    % legend image code/.code={%
    % \draw[#1, draw=none] (0cm,-0.1cm) rectangle (0.6cm,0.1cm);}
]

% \addplot[line width=1pt, GreenColor,mark=*] coordinates
% {(1, 51.2) (2, 47.7) (3, 50.5) (4, 62.2) (5, 63.0)(6, 75.2)};
% Read
\addplot[line width=1pt, GreenColor, densely dotted]
table[x=x,y=y] {
x y
0 0.6525 
1 0.7245 
2 0.8033 
3 0.8767 
4 0.9334 
5 0.9690 
6 0.9849 
7 0.9925 
8 0.9963 
};
\addplot[line width=1pt, AcornOneColor, densely dashed]
table[x=x,y=y] {
x y
0 0.5983
1 0.6708
2 0.7496
3 0.8296
4 0.8989 
5 0.9488
6 0.978
7 0.9921
8 0.9962 
};
\addplot[line width=1pt, AcornGammaColor, densely dashed]
table[x=x,y=y] {
x y
0 0.5506 
1 0.6059
2 0.6735 
3 0.7464 
4 0.8231 
5 0.8865 
6 0.9359 
7 0.9693 
8 0.9873 
};
% ## 4times

\end{axis}
\end{tikzpicture}
\vspace{-3mm}
\caption{nprobe vs. recall}
\label{fig:exp_spann_lower_replica_metrics_nprobe}
\end{subfigure}
\vspace{-1mm}
\caption{\spann on \gist: reducing replication decreases posting list size, but does not necessarily reduce data read per query at the same recall (left) and can negatively impact QPS-recall trade-off (middle) due to lower-quality indexing requiring higher \texttt{nprobe} values to reach the same recall (right).
}
\vspace{-3mm}
\label{fig:exp_spann_lower_replica_metrics}
\end{figure}

\paragraph{\spann: Replication Count Impacts Index Quality}
Another method for reducing \spann's posting list sizes without increasing posting list count (i.e., BKT tree costs) is to decrease vector replication. We test two \spann indexes with reduced vector replication counts of 4 and 2 in \cref{fig:exp_spann_lower_replica_metrics}: Despite data read per query being reduced at various \texttt{nprobe} values due to smaller posting list sizes (\cref{tbl:spann_index_size}), these indexes do not necessarily result in better QPS-recall trade-offs: reducing replication count also reduces index quality, which results in a higher \texttt{nprobe} value required to reach the same recall---for example, the replica=2 index requires 3-4$\times$ higher \texttt{nprobe} versus the replica=8 index to reach the same recall at all recall levels (\cref{fig:exp_spann_lower_replica_metrics_nprobe}), ultimately resulting in \textit{more} data read per query at the same recall (\cref{fig:exp_spann_lower_replica_metrics_dataread}) and lower QPS (e,g, 2.00$\times$ more data read and 1.92$\times$ lower QPS vs. replica=8 @ 0.97 recall).
% Yet, when caching is involved, low-replication indexes's smaller posting lists improve cache-friendliness, which we explore in detail in \cref{sec:exp_caching}.

% \subsubsection{Tuning \diskann}
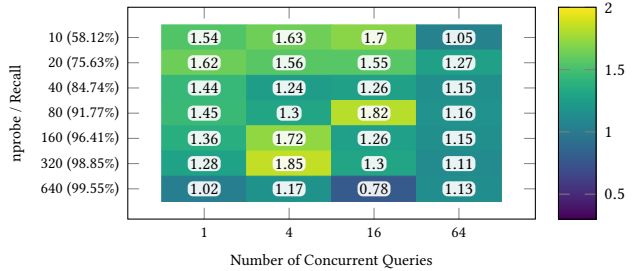
\begin{figure}[t]\captionsetup[subfigure]{font=footnotesize}
\pgfplotsset{scaled y ticks=false}
\centering

\begin{subfigure}[b]{\linewidth}
\begin{tikzpicture}
\begin{axis}[
    width=7cm, height=4.4cm,
    matrix plot,
    colormap/viridis,
    xlabel={Number of Concurrent Queries},
    ylabel={nprobe / Recall},
    xtick={1,2,3,4},
    xticklabels={1, 4, 16, 64},
    ytick={1,2,3,4,5,6,7},
    yticklabels={
        10 (58.12\%), 
        20 (75.63\%), 
        40 (84.74\%), 
        80 (91.77\%), 
        160 (96.41\%), 
        320 (98.85\%), 
        640 (99.55\%), 
    },
    label style={font=\scriptsize},
    tick label style={font=\scriptsize},
    colorbar,
    point meta min=0.3,
    point meta max=2.0,
    nodes near coords,
    nodes near coords align={center},
    every node near coord/.append style={font=\scriptsize, color=black, opacity=0.9, fill=white, text opacity=1, rounded corners=2pt, inner sep=0.8pt},
    y dir=reverse,
    ]
\addplot [matrix plot*, point meta=explicit] table [meta=ratio] {
x y ratio
1 1 1.540372671
2 1 1.626946513
3 1 1.703565526
4 1 1.05472834

1 2 1.62
2 2 1.557500538
3 2 1.552652402
4 2 1.266473836

1 3 1.436813187
2 3 1.243666116
3 3 1.26342574
4 3 1.145469435

1 4 1.451361868
2 4 1.300985595
3 4 1.817148413
4 4 1.160802856

1 5 1.362318841
2 5 1.722874863
3 5 1.261082182
4 5 1.151615603

1 6 1.278481013
2 6 1.847285642
3 6 1.300357649
4 6 1.109772423

1 7 1.022222222			
2 7 1.165075669
3 7 0.776879226
4 7 1.12763596	

};
\end{axis}
\end{tikzpicture}
\vspace{-1.5mm}
\caption{Ratio of (R=256 QPS/R=64 QPS) of \diskann on \gist}
\end{subfigure}
\vspace{-5mm}
\caption{QPS ratio of the alternative \diskann index with denser graph (R=256) over the default graph with R=64. The denser graph achieves consistent QPS increases across all recalls and concurrencies on remote storage.
}\vspace{-3mm}
\label{fig:exp_diskann_denser_graph}
\end{figure}
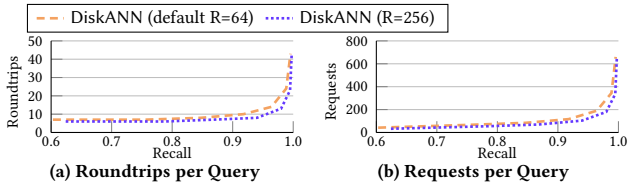
\begin{figure}[t]\captionsetup[subfigure]{font=footnotesize}
\pgfplotsset{scaled y ticks=false}
\centering
\begin{subfigure}[b]{0.49\linewidth}
\begin{tikzpicture}

\begin{axis}[
    xtick=data,
    width=48mm,
    height=28mm,
    ymin=0,
    ymax=50,
    axis y line*=none,
    axis x line*=none,
    ytick={0, 10, 20, 30, 40, 50},
    yticklabels={0, 10, 20, 30, 40, 50},
    xlabel=Recall,
    xlabel style={yshift = 1.5ex},
    label style={font=\scriptsize},
    ylabel style={yshift=-1ex,xshift=-0.5ex, font=\scriptsize},
    xmin = 0.6,
    xmax = 1,
    xtick = {0.6, 0.7, 0.8, 0.9, 1.0},
    xticklabels = {0.6, 0.7, 0.8, 0.9, 1.0},
    tick label style={font=\scriptsize},
    x tick label style={yshift=0.5ex},
    legend style={
        at={(-0.1,1.09)},anchor=south west,column sep=0pt,
        legend image post style={xscale=0.6},
        row sep = -0.4pt,
        draw=black,fill=white,
        inner ysep=0.1pt,
        /tikz/every even column/.append style={column sep=2pt},
        font=\footnotesize
    },
    legend cell align={left},
    legend columns=4,
    ylabel={Roundtrips},
    ymajorgrids,
    every axis plot/.append style={thick}
    % legend image code/.code={%
    % \draw[#1, draw=none] (0cm,-0.1cm) rectangle (0.6cm,0.1cm);}
]

% \addplot[line width=1pt, GreenColor,mark=*] coordinates
% {(1, 51.2) (2, 47.7) (3, 50.5) (4, 62.2) (5, 63.0)(6, 75.2)};
% Read
\addplot[line width=1pt, HeuristicColor, densely dashed]
table[x=x,y=y] {
x y
0.5812  7
0.7563  7
0.8474  8
0.9177  10
0.9641  14
0.9885  24
0.9955  43
};
\addlegendentry{\diskann (default R=64)}
\addplot[line width=1pt, DiskAnnColor, densely dotted]
table[x=x,y=y] {
x y
0.6241  6
0.7874  6
0.8681  7
0.9400  8
0.9799  13
0.9948  23
0.9974  42
};
\addlegendentry{\diskann (R=256)}
% ## 4times

\end{axis}
\end{tikzpicture}
\vspace{-6.5mm}
\caption{Roundtrips per Query}
\label{fig:exp_diskann_denser_graph_metrics_roundtrips}
\end{subfigure}
\begin{subfigure}[b]{0.49\linewidth}
\begin{tikzpicture}

\begin{axis}[
    xtick=data,
    width=48mm,
    height=28mm,
    ymin=0,
    ymax=800,
    axis y line*=none,
    axis x line*=none,
    ytick={0, 200, 400, 600, 800},
    yticklabels={0, 200, 400, 600, 800},
    xlabel=Recall,
    xlabel style={yshift = 1.5ex},
    label style={font=\scriptsize},
    ylabel style={yshift=-1ex,xshift=-0.5ex, font=\scriptsize},
    xmin = 0.6,
    xmax = 1,
    xtick = {0.6, 0.7, 0.8, 0.9, 1.0},
    xticklabels = {0.6, 0.7, 0.8, 0.9, 1.0},
    tick label style={font=\scriptsize},
    x tick label style={yshift=0.5ex},
    legend style={
        at={(-0.4,1.09)},anchor=south west,column sep=0pt,
        legend image post style={xscale=0.6},
        row sep = -0.4pt,
        draw=black,fill=white,
        inner ysep=0.1pt,
        /tikz/every even column/.append style={column sep=2pt},
        font=\footnotesize
    },
    legend cell align={left},
    legend columns=4,
    ylabel={Requests},
    ymajorgrids,
    every axis plot/.append style={thick}
    % legend image code/.code={%
    % \draw[#1, draw=none] (0cm,-0.1cm) rectangle (0.6cm,0.1cm);}
]

% \addplot[line width=1pt, GreenColor,mark=*] coordinates
% {(1, 51.2) (2, 47.7) (3, 50.5) (4, 62.2) (5, 63.0)(6, 75.2)};
% Read
\addplot[line width=1pt, HeuristicColor, densely dashed]
table[x=x,y=y] {
x y
0.5812  38.45
0.7563  67.19
0.8474  84.58
0.9177  116.56
0.9641  189.85
0.9885  344.5
0.9955  660.16
};
\addplot[line width=1pt, DiskAnnColor, densely dotted]
table[x=x,y=y] {
x y
0.6241  32.45
0.7874  55.04
0.8681  69.63
0.9400  103.72
0.9799  179.55
0.9948  336.27
0.9974  653.06
};
% ## 4times

\end{axis}
\end{tikzpicture}
\vspace{-6.5mm}
\caption{Requests per Query}
\label{fig:exp_diskann_denser_graph_metrics_requests}
\end{subfigure}

\vspace{-2mm}
\caption{\diskann on \gist: a denser graph reduces the number of roundtrips and requests to storage per query, benefitting on-cloud performance.
}
\vspace{-3mm}
\label{fig:exp_diskann_denser_graph_metrics}
\end{figure}
\paragraph{\diskann: Denser Graphs Benefit Cloud Setups}
\Cref{fig:exp_diskann_denser_graph} reports the QPS ratio of an alternative \diskann index built with a higher density (R=256) versus the default \diskann index built with R=64. While having a larger size (8.2GB vs. 7.1GB), the former achieves consistent (up to 1.85$\times$) QPS gains versus the latter on all scenarios. The detailed metrics in \cref{fig:exp_diskann_denser_graph_metrics} showcase reasons: the R=256 graph reduces the number of roundtrips to remote storage (\cref{fig:exp_diskann_denser_graph_metrics_roundtrips}) and the number of requests per query (\cref{fig:exp_diskann_denser_graph_metrics_requests}) at all recalls, which address \diskann's I/O bottleneck (\cref{tbl:diskann_index_tuning}) to improve query performance.

% However, this performance improvement comes at the cost of incurring significantly more distance computations performed on the product-quantized vectors which \diskann loads into memory before query serving (\textcolor{red}{cref}). The original \diskann paper recommends R values in the range of $[30, 150]$, which is because the index was orignially designed for on-disk querying~\cite{jayaram2019diskann}; accordingly, we can see that the denser R=256 graph significantly underperforms versus the default R=64 graph on-disk (\cref{fig:exp_diskann_denser_graph_metrics_qps}), as the Disk I/O saved does not make up for the large increase in distance computations made on PQed vectors (\cref{tbl:exp_diskann_denser_graph_metrics_breakdown}).

\input{sections/plots/exp_diskann_larger_beamwidth}
\paragraph{\diskann: High Beamwidth for High Recall Ad-Hoc Queries}
Alternatively, a higher search beamwidth may also reduce the number of roundtrips per query, which we study in \cref{fig:exp_diskann_larger_beamwidth}. \Cref{fig:exp_diskann_larger_beamwidth_recall_roundtrips} shows that there are only significant reductions at high recalls, from 43 to 14 at 0.995 recall, which consequently improves QPS by 1.85$\times$ with concurrency=4 (\cref{fig:exp_diskann_larger_beamwidth_recall_qps}). However, higher beamwidths incur more requests per query (\cref{fig:exp_diskann_larger_beamwidth_concurrency_requests}), which decreases QPS under all other scenarios: each query still incurs 7 roundtrips regardless of beamwidth at 0.8 recall, leading to a 1.83$\times$ QPS decrease in correspondance to the 1.71$\times$ request increase from beamwidth=16 to 64.
On the other hand, higher beamwidths saturates the IOPS limit more easily at high concurrencies: while I/O bandwidth to remote storage is far from saturated with beamwidth=64 even at concurrency=64 (83.4 MB/s, \cref{fig:exp_diskann_larger_beamwidth_concurrency_requests_bw}), the 20K IOPS limit has been reached even with beamwidth=32 (\cref{fig:exp_diskann_larger_beamwidth_concurrency_requests}), resulting in beamwidth=32 having 1.10$\times$ QPS versus beamwidth=64 at concurrency=64 (\cref{fig:exp_diskann_larger_beamwidth_concurrency_qps}).

\begin{tcolorbox}[colback=gray!10,colframe=gray!40,boxrule=0.5pt,arc=2pt,left=5pt,right=5pt,top=5pt,bottom=5pt]
\paragraph{A2: How to design indexes?}
\spann and \diskann both have parameters which can address their specific bottlenecks for cloud-native vector search: For \spann, this would be using more centroids for high recall/concurrency scenarios to trade lower data read per query for BKT tree computations; for \diskann, this would be using a denser graph to trade fewer roundtrips for more distance computations, while selectively using higher beamwidth for high-recall setups to further reduce roundtrips when the concurrency is sufficiently low such that the higher requests per query does not saturate IOPS limit.
\end{tcolorbox}

\input{sections/plots/exp_caching_e2e}
\begin{figure}[t]\captionsetup[subfigure]{font=footnotesize}
\pgfplotsset{scaled y ticks=false}
\centering
\begin{subfigure}[b]{0.49\linewidth}
\begin{tikzpicture}

\begin{axis}[
    xtick=data,
    width=\figwidth,
    height=28mm,
    ymin=0,
    ymax=1,
    axis y line*=none,
    axis x line*=none,
    ytick={0, 0.2, 0.4, 0.6, 0.8, 1},
    yticklabels={0, 0.2, 0.4, 0.6, 0.8, 1},
    xlabel=Recall,
    xlabel style={yshift = 1.5ex},
    label style={font=\scriptsize},
    ylabel style={yshift=-1ex,xshift=-0.5ex, font=\scriptsize},
    xmin = 0.6,
    xmax = 1,
    xtick = {0.6, 0.7, 0.8, 0.9, 1.0},
    xticklabels = {0.6, 0.7, 0.8, 0.9, 1.0},
    tick label style={font=\scriptsize},
    x tick label style={yshift=0.5ex},
    legend style={
        at={(0,1.09)},anchor=south west,column sep=0pt,
        legend image post style={xscale=0.6},
        row sep = -0.4pt,
        draw=black,fill=white,
        inner ysep=0.1pt,
        /tikz/every even column/.append style={column sep=2pt},
        font=\footnotesize
    },
    legend cell align={left},
    legend columns=4,
    ylabel={Cache hit rate},
    ymajorgrids,
    every axis plot/.append style={thick}
    % legend image code/.code={%
    % \draw[#1, draw=none] (0cm,-0.1cm) rectangle (0.6cm,0.1cm);}
]

% \addplot[line width=1pt, GreenColor,mark=*] coordinates
% {(1, 51.2) (2, 47.7) (3, 50.5) (4, 62.2) (5, 63.0)(6, 75.2)};
% Read
\addplot[line width=1pt, OursColor, densely dashed]
table[x=x,y=y] {
x y
0.6525  0.1053
0.7245  0.1315
0.8033  0.1506
0.8767  0.1632
0.9334  0.1774
0.9690  0.1875
0.9849  0.1909
0.9925  0.1869
0.9963  0.1772
};
\addlegendentry{1GB cache}
% \addplot[line width=1pt, AcornOneColor, densely dashed]
% table[x=x,y=y] {
% x y
% 0.6525 0.1066
% 0.7245 0.1598
% 0.8033 0.2164
% 0.8767 0.2523
% 0.9334 0.2723
% 0.9690 0.292
% 0.9849 0.3104
% 0.9925 0.3155
% 0.9963 0.3098
% };
% \addlegendentry{2GB cache}
\addplot[line width=1pt, AcornGammaColor, densely dashed]
table[x=x,y=y] {
x y
0.6525  0.1057
0.7245  0.1614
0.8033  0.2462
0.8767  0.3519
0.9334  0.4218
0.9690  0.4665
0.9849  0.4907
0.9925  0.5142
0.9963  0.5271
};
\addlegendentry{4GB cache}
\addplot[line width=1pt, HeuristicColor, densely dashed]
table[x=x,y=y] {
x y
0.6525  0.1064
0.7245  0.1609
0.8033  0.2464
0.8767  0.3642
0.9334  0.5068
0.9690  0.6466
0.9849  0.7369
0.9925  0.7863
0.9963  0.8126
};
\addlegendentry{8GB cache}
\end{axis}
\end{tikzpicture}
\vspace{-6.5mm}
\caption{\spann}
\label{fig:exp_caching_hitrate_spann}
\end{subfigure}
\begin{subfigure}[b]{0.49\linewidth}
\begin{tikzpicture}

\begin{axis}[
    xtick=data,
    width=\figwidth,
height=28mm,
    ymin=0,
    ymax=1,
    axis y line*=none,
    axis x line*=none,
    ytick={0, 0.2, 0.4, 0.6, 0.8, 1},
    yticklabels={0, 0.2, 0.4, 0.6, 0.8, 1},
    xlabel=Recall,
    xlabel style={yshift = 1.5ex},
    label style={font=\scriptsize},
        ylabel style={yshift=-1ex,xshift=-0.5ex, font=\scriptsize},
    xmin = 0.6,
    xmax = 1,
    xtick = {0.6, 0.7, 0.8, 0.9, 1.0},
    xticklabels = {0.6, 0.7, 0.8, 0.9, 1.0},
    x tick label style={yshift=0.5ex},
    tick label style={font=\scriptsize},
    legend style={
        at={(-0.2,1.1)},anchor=south west,column sep=2pt,
        draw=black,fill=white,
        inner ysep=0.1pt,
        /tikz/every even column/.append style={column sep=5pt},
        font=\footnotesize
    },
    legend cell align={left},
    legend columns=5,
    ylabel={Cache hit rate},
    ymajorgrids,
    every axis plot/.append style={thick}
    % legend image code/.code={%
    % \draw[#1, draw=none] (0cm,-0.1cm) rectangle (0.6cm,0.1cm);}
]

% \addplot[line width=1pt, GreenColor,mark=*] coordinates
% {(1, 51.2) (2, 47.7) (3, 50.5) (4, 62.2) (5, 63.0)(6, 75.2)};
% Read
\addplot[line width=1pt, OursColor, densely dashed]
table[x=x,y=y] {
x y
0.5812  0.3624
0.7563  0.3823
0.8474  0.3634
0.9177  0.3477
0.9641  0.3459
0.9885  0.3499
0.9955  0.3507
};
\addplot[line width=1pt, AcornGammaColor, densely dashed]
table[x=x,y=y] {
x y
0.5812  0.3624
0.7563  0.3822
0.8474  0.3634
0.9177  0.3477
0.9641  0.3461
0.9885  0.3914
0.9955  0.489
};
% \addplot[line width=1pt, AcornGammaColor, densely dashed]
% table[x=x,y=y] {
% x y
% 0.6525  515.7195
% 0.8033  277.9886
% 0.9334  133.8979
% 0.9690  61.7577
% 0.9849  47.7292
% 0.9925  60.2645
% 0.9963  35.6961
% };
\end{axis}
\end{tikzpicture}
\vspace{-6.5mm}
\caption{\diskann}
\label{fig:exp_caching_hitrate_diskann}
\end{subfigure}
\vspace{-1mm}
\caption{Cache hit rate versus size of \spann and \diskann on \gist, which both exhibit higher hit rate at higher recalls.
}
\vspace{-3mm}
\label{fig:exp_caching_hitrate}
\end{figure}
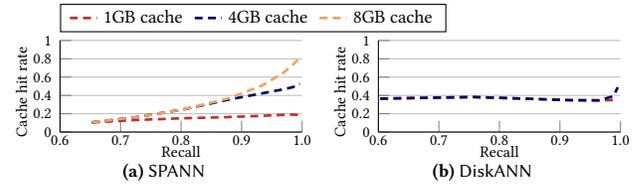
\subsection{How to Utilize Caching?}
\label{sec:exp_caching}
This section studies how to effectively utilize local caching for index segments in cloud-native vector search. We evaluate caching gains with various cache sizes set up following \cref{sec:exp_setup}, and investigate performance gains and metrics under different index parameterizations and datasets to understand how to effectively utilize caching for on-remote storage querying. We perform all experiments with a cold start, i.e., empty cache, in this section. We report caching gains on \gist in \cref{fig:exp_caching_e2e},\footnote{All 1K queries on \gist read less than 4GB data combined with \diskann, making 4GB and 8GB caches functionally identical. We omit the latter from the plot for brevity.} and hitrate versus recall in \cref{fig:exp_caching_hitrate}. 

\input{sections/plots/exp_caching_spann_detailed}
\paragraph{\spann: Cache Hits saves Network Resources}
\spann benefits more from caching at high concurrencies, larger cache sizes, and higher recalls. The first two factors are straightforward: \spann exhibits more I/O congestion at high concurrencies which caching addresses, and a larger cache holds more posting lists for higher hit rate (\cref{fig:exp_caching_hitrate_spann}), which is especially noticeable at high recalls: \spann's large amount of data read per query at high recalls (\cref{fig:exp_index_metrics_dataread}) leads to high cache locality, with queries reading more overlapping posting lists; this leads to the 1GB cache being insufficient for holding all 'hot' posting lists: while the 4GB cache's hit rate reaches 0.68 at 0.995 recall, the 1GB cache's hit rate in fact drops from 0.19 to 0.17 from 0.98 to 0.995 recall, explaining the significant performance difference between the 1GB and 4GB caches at this recall (\cref{fig:exp_caching_e2e}).
In particular, \spann's cache hits benefits both individual queries and workloads as a whole (\cref{fig:exp_caching_spann_detailed}; for the former, it reduces the bandwidth pressure of concurrently fetching posting lists, which we observe for an example query on \gist where the significant queuing of posting list requests (\cref{fig:exp_caching_spann_no_hit}) is reduced with a 4GB cache (\cref{fig:exp_caching_spann_some_hits})---the I/O latencies exhibit a bimodal distribution between cache hits and misses. Caching reduces the total data read and number of requests of the workload for the latter, which, in addition to addressing bandwidth congestion, similarly addresses IOPS congestion as we have observed on the \deep dataset (\cref{fig:exp_spann_deep_getqps}) at high recalls and concurrencies (\cref{fig:exp_caching_spann_deep}).

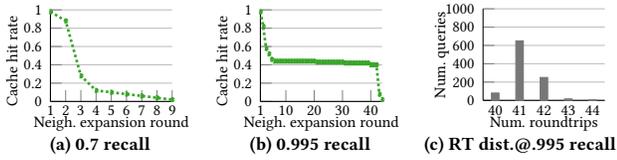
\begin{figure}[t]\captionsetup[subfigure]{font=footnotesize}
\pgfplotsset{scaled y ticks=false}
\centering
\begin{subfigure}[b]{0.32\linewidth}
\begin{tikzpicture}

\begin{axis}[
    xtick=data,
    width=32mm,
    height=28mm,
    ymin=0,
    ymax=1,
    axis y line*=none,
    axis x line*=none,
    ytick={0, 0.2, 0.4, 0.6, 0.8, 1},
    yticklabels={0, 0.2, 0.4, 0.6, 0.8, 1},
    xlabel=Neigh. expansion round,
    xlabel style={yshift = 1.5ex},
    label style={font=\scriptsize},
    ylabel style={yshift=-1ex,xshift=-0.5ex, font=\scriptsize},
    xmin = 1,
    xmax = 9,
    xtick = {1, 2, 3, 4, 5, 6, 7, 8, 9},
    xticklabels = {1, 2, 3, 4, 5, 6, 7, 8, 9},
    tick label style={font=\scriptsize},
    x tick label style={yshift=0.5ex},
    legend style={
        at={(-0.6,1.09)},anchor=south west,column sep=0pt,
        legend image post style={xscale=0.6},
        row sep = -0.4pt,
        draw=black,fill=white,
        inner ysep=0.1pt,
        /tikz/every even column/.append style={column sep=2pt},
        font=\footnotesize
    },
    legend cell align={left},
    legend columns=4,
    ylabel={Cache hit rate},
    ymajorgrids,
    every axis plot/.append style={thick}
    % legend image code/.code={%
    % \draw[#1, draw=none] (0cm,-0.1cm) rectangle (0.6cm,0.1cm);}
]

% \addplot[line width=1pt, GreenColor,mark=*] coordinates
% {(1, 51.2) (2, 47.7) (3, 50.5) (4, 62.2) (5, 63.0)(6, 75.2)};
% Read
\addplot[line width=1pt, GreenColor, densely dotted, mark = *, mark size=0.5pt]
table[x=x,y=y] {
x y
1 0.98
2 0.88
3 0.28
4 0.12
5 0.10
6 0.08
7 0.06
8 0.04
9 0.02
};
% ## 4times

\end{axis}
\end{tikzpicture}
\vspace{-2.5mm}
\caption{0.7 recall}
\label{fig:exp_diskann_neigh_expansion_low_recall}
\end{subfigure}
\begin{subfigure}[b]{0.32\linewidth}
\begin{tikzpicture}

\begin{axis}[
    xtick=data,
    width=32mm,
    height=28mm,
        ymin=0,
    ymax=1,
    axis y line*=none,
    axis x line*=none,
    ytick={0, 0.2, 0.4, 0.6, 0.8, 1},
    yticklabels={0, 0.2, 0.4, 0.6, 0.8, 1},
    xlabel=Neigh. expansion round,
    xlabel style={yshift = 1.5ex},
    label style={font=\scriptsize},
    ylabel style={yshift=-1ex,xshift=-0.5ex, font=\scriptsize},
    xmin = 1,
    xmax = 44,
    xtick = {1, 10, 20, 30, 40},
    xticklabels = {1, 10, 20, 30, 40},
    tick label style={font=\scriptsize},
    x tick label style={yshift=0.5ex},
    legend style={
        at={(-0.4,1.09)},anchor=south west,column sep=0pt,
        legend image post style={xscale=0.6},
        row sep = -0.4pt,
        draw=black,fill=white,
        inner ysep=0.1pt,
        /tikz/every even column/.append style={column sep=2pt},
        font=\footnotesize
    },
    legend cell align={left},
    legend columns=4,
    ylabel={Cache hit rate},
    ymajorgrids,
    every axis plot/.append style={thick}
    % legend image code/.code={%
    % \draw[#1, draw=none] (0cm,-0.1cm) rectangle (0.6cm,0.1cm);}
]

% \addplot[line width=1pt, GreenColor,mark=*] coordinates
% {(1, 51.2) (2, 47.7) (3, 50.5) (4, 62.2) (5, 63.0)(6, 75.2)};
% Read
\addplot[line width=1pt, GreenColor, densely dotted, mark = *, mark size=0.5pt]
table[x=x,y=y] {
x y
1 0.98
2 0.82
3 0.58
4 0.52
5 0.46
6 0.44
7 0.44
8 0.44
9 0.44
10 0.44
11 0.44
12 0.44
13 0.44
14 0.44
15 0.44
16 0.44
17 0.44
18 0.44
19 0.44
20 0.44
21 0.43
22 0.43
23 0.43
24 0.43
25 0.43
26 0.43
27 0.43
28 0.43
29 0.43
30 0.43
31 0.42
32 0.42
33 0.42
34 0.42
35 0.42
36 0.42
37 0.42
38 0.42
39 0.42
40 0.40
41 0.40
42 0.40
43 0.08
44 0.02
};
% ## 4times

\end{axis}
\end{tikzpicture}
\vspace{-2.5mm}
\caption{0.995 recall}
\label{fig:exp_diskann_neigh_expansion_high_recall}
\end{subfigure}
\begin{subfigure}[b]{0.32\linewidth}
\centering
\begin{tikzpicture}

\begin{axis}[
    ybar,
    clip=false,
    xlabel style={yshift = 1.5ex},
    width=32mm,
    height=28mm,
    bar width=1mm,
    ymin=0,
    ymax=1000,
    ylabel style={yshift = -2ex},
    axis y line*=none,
    axis x line*=none,
    ytick={0, 200, 400, 600, 800, 1000},
    yticklabels={0, 200, 400, 600, 800, 1000},
    xtick={0, 1, 2, 3, 4},
    xticklabels={40, 41, 42, 43, 44},
    x tick label style={yshift = 1ex},
    xmin=-0.5,
    xmax = 4.5,
        xtick style ={draw=none},
        x tick label style={yshift = 0.5ex},
    ymajorgrids,
    tick label style={font=\scriptsize},
    legend style={
        font=\scriptsize,
        /tikz/every even column/.append style={column sep=0.5cm},
        legend columns = 3,
        at={(-0.15, 1.1)}, anchor=south west
    },
    label style={font=\scriptsize},
    ylabel={Num. queries},
    xlabel={Num. roundtrips},
    area legend
    ]

      \addplot[LRUColor,fill=LRUColor] coordinates {(0,80) (1,650) (2,250) (3,16) (4,4)};

    % \node[anchor=south west, red, rotate=90,
    %     font=\footnotesize\bfseries] 
    %     at (axis cs: 3.2, 0) 
    %     {100\% failure};

\end{axis}
    %\node[draw,fill=white, align=left] at (3.3, 1.8) {775s};
    
    % \draw[fill=white,draw=white] (3.0,1.4) -- (3.6,1.6) -- (3.6,1.8) -- (3.0,1.6) -- cycle;
    % \draw[draw=black] (3.0,1.6) -- (3.6,1.8);
    % \draw[draw=black] (3.0,1.4) -- (3.6,1.6);
    
\end{tikzpicture}
\vspace{-2.5mm}
\caption{RT dist.@.995 recall}
\label{fig:exp_diskann_neigh_expansion_dist}
\end{subfigure}

\vspace{-1mm}
\caption{\diskann's per-neighbor expansion round cache hitrate: the first few rounds corresponding to expanding around the graph's entry point exhibits the highest hitrates regardless of query parameterization.
}
\vspace{-3mm}
\label{fig:caching_diskann_detailed}
\end{figure}
\paragraph{\diskann: Cache Hits Benefit Early Expansion Latency}  \diskann receives similar benefits from the 1GB and 4GB caches under low concurrencies (up to 1.41$\times$ for both sizes for concurrency=4), and unlike \spann, noticeably benefits more from caching at low recalls (1.41$\times$@0.7 recall vs. 1.02$\times$@0.99 recall w/ 4GB cache, \cref{fig:exp_caching_e2e}) \textit{despite} higher cache hit rates from better cache locality due to overlapping query blocks (0.49 vs. 0.35 hitrate from 0.7 to 0.99 recall w/ 4GB cache). These factors can be attributed to \diskann's unique interaction between beamwidth and cache hits (\cref{sec:caching_benefit}): \Cref{fig:caching_diskann_detailed} shows that the first few roundtrips (i.e., the \diskann graph's entry point neighborhood) consistently have near-1 hitrates regardless of recall: these saved initial roundtrips comprise a larger portion of traversal at low recalls (\cref{fig:exp_diskann_neigh_expansion_low_recall}) versus at high recalls (\cref{fig:exp_diskann_neigh_expansion_high_recall}). Interestingly, we also observe a trimodal distribution in per-expansion cache hit rate in \cref{fig:exp_diskann_neigh_expansion_high_recall}, where the hit rate of the latest rounds (>42) performed by long-tailed, difficult queries (\cref{fig:exp_diskann_neigh_expansion_dist}) are near-0, as these rounds require visiting seldomly re-visited sparse graph regions~\cite{jayaram2019diskann}. 
However, like \spann, under IOPS-saturated settings (e.g., concurrency=64 and >0.99 recall), \diskann does gain performance with larger cache size, with the 4GB cache achieving 1.45$\times$ QPS increase versus 1GB cache's 1.30$\times$ (\cref{fig:exp_diskann_cache_concurrency64}): while cache hits may still not reduce the number of roundtrips, they effectively reduce the number of requests per query and consequently IOPS.

\input{sections/plots/exp_spann_cache_tuning}
% \subsection{Cache-Aware Indexing}
% \label{sec:exp_detailed}
% This section studies how caching interacts with various indexing configurations. Building upon our findings in \cref{sec:exp_caching}, we investigate how \spann and \diskann indexes can be tuned to extract more benefits out of caching.

\paragraph{\spann: Reducing Replication Increases Hit Rate}
One potential drawback of \spann's vector replication is that it increases posting list size (\cref{tbl:spann_index_size}), which reduces the number of posting lists a fixed-size cache holds and potentially the cache hit rate (\cref{sec:caching_benefit}). We observe this in \cref{fig:exp_spann_cache_tuning} where we evaluate the various low-replication \spann indexes under various cache sizes:
while low-replication indexes reduce QPS-recall trade-off without a cache (\cref{fig:exp_spann_lower_replica_metrics_qps}), when combined with a sufficiently-sized cache (i.e., 4GB), their smaller posting lists significantly increases the cache hit rate, and consequently the data read per query. For example, despite the replica=2 index requiring \texttt{nprobe=1024} to reach 0.97 recall versus the replica=8 index requring \texttt{nprobe=256}, it still reads 3.09$\times$ less data per query (\cref{fig:exp_spann_cache_tuning_4gb_dataread}) due to its significantly higher cache hit rate (0.778 vs. 0.467, \cref{fig:exp_spann_cache_tuning_4gb_hitrate}), leading to a 2.03$\times$ QPS increase at this recall and concurrency=4 (\cref{fig:exp_spann_cache_tuning_4gb_qps}). Interestingly, with smaller (e.g., 1GB, \cref{fig:exp_spann_cache_tuning_1gb}) or even larger (e.g., 8GB \cref{fig:exp_spann_cache_tuning_8gb}) cache size, replica=8 becomes optimal again: low-replcation indexes do not achieve sufficiently higher cache hit rate for the former (\cref{fig:exp_spann_cache_tuning_1gb_hitrate}), and replica=8 already has sufficiently high hitrate for the latter (\cref{fig:exp_spann_cache_tuning_8gb_hitrate}).

% A key factor limiting \spann's benefit from caching is its large amount of data read per query relative to the cache size resulting in low cache hit rates with small cache sizes---a small cache is often insufficient for holding all hot posting lists at high-recall querying configurations (\cref{fig:exp_caching_hitrate_spann}). One effective method for reducing posting list sizes is to reduce the replication factor (\cref{sec:exp_parameterization}); while this approach can negatively impact QPS-recall trade-off without a cache due to significant recall drops at the same \texttt{nprobe} value from lower-quality posting lists (\cref{fig:exp_spann_lower_replica_metrics}), it shows promising gains when combined with caching (\cref{fig:exp_spann_cache_tuning}): with a 4GB cache, the \textit{no replica} \spann index (equivalent of building a BKT tree over an IVF index) exhibits up to 1.18$\times$ higher QPS at all recalls versus the default replica=8 index (\cref{fig:exp_spann_cache_tuning_qps}); this is because the no replica index has a size of only \textbf{4.1GB} with 60KB posting lists resulting in the 4GB cache being able to hold almost all of the index, reaching 0.988 cache hit rate (\cref{fig:exp_spann_cache_tuning_hitrate}) at 0.995 recall. Therefore, despite it needing \texttt{nprobe}=16384 to reach 0.995 recall versus the default replica=8 index needing \texttt{nprobe}=2048, it still reads 76.4$\times$ less data per query due to the cache hits.

% \input{sections/plots/exp_caching_warm}
\input{sections/plots/exp_diskann_cache_tuning}
\paragraph{\diskann: Caching Enables Higher Beamwidth Usage}
We have observed caching to bring negligible benefits to the latter roundtrips of queries due to low hitrates (\cref{fig:caching_diskann_detailed}); in fact, for ad-hoc queries (i.e., concurrency=1), the higher the beamwidth, the lower the QPS benefits due to the decreased possibility of caching all nodes of a roundtrip---for example, it is much more likely for a roundtrip to be saved via caching under W=4 (\cref{sec:caching_benefit}) versus W=16 (\cref{fig:diskann_cache_teaser_roundtrips}). However, under cases where high beamwidth is beneficial (e.g., high-recall, ad-hoc querying as shown in \cref{fig:exp_diskann_cache_gains_latency}), caching benefits (1.25$\times$ at W=4) are insignificant versus gains from increasing beamwidth (7.58$\times$ from W=4 to 64), hence the latter should be used regardless.

\begin{tcolorbox}[colback=gray!10,colframe=gray!40,boxrule=0.5pt,arc=2pt,left=5pt,right=5pt,top=5pt,bottom=5pt]
\paragraph{A3: How to Utilize Caching?}
\spann benefits from caching as cache hits reduce I/O bandwidth and/or IOPS pressure of concurrent posting list fetching. Benefits can be increased under moderate cache sizes (i.e., insufficient for holding enough hot posting lists under high replication) by reducing replication factor to reduce posting list size and increase cache hit rate; effectiveness of this approach diminishes when the cache is too small even when replication is reduced regardless, or when the cache is large enough.

\diskann's multi-node expansions result in its individual queries only significantly benefitting from caching latency-wise in its first few expansions; hence, for non IOPS-saturated settings, it may be sufficient to use a fixed-size cache which holds the neighborhood around the graph's entry point as suggested in the original \diskann paper~\cite{jayaram2019diskann}. A standard (e.g., LRU) cache may still be used under IOPS-saturated settings. While high beamwidth decreases caching gains due to further decreased chances of saving roundtrips, caching gains are often insignificant versus beamwidth-induced gains; beamwidth should be set following \textbf{A2} (\cref{sec:exp_parameterization}) as if there is no cache.
\end{tcolorbox}

\section{Related Work}
\label{sec:related}
\paragraph{Cloud-Native Vector Search}
TurboPufer and Amazon S3 Vector have recently began offering cloud-native vector search services~\cite{turbopuffer, amazons3vector}. Differing from prior providers utilizing cloud storage only to persist vector indexes and perform querying entirely in memory~\cite{opensearch, guo2022manu}, these cloud-native providers query vector indexes directly on cloud storage, fetching index segments on demand. Notably, cloud-native providers have only explored querying with cluster indexes, and only utilize local caching to hold index metadata (e.g., SPANN BKT trees, \cref{sec:caching_background}) of hot indexes. Our work compliments development of these cloud-native vector search systems by identifying bottlenecks of the two index classes on remote storage, and accordingly providing suggestions on the index (and parameters) and caching strategy, to use for different workloads (\cref{sec:experiments}).

\paragraph{Data placement for on-disk graph indexes}
There exists works aimed at optimizing the node placement of on-disk graph-based indexes aiming to improve data locality~\cite{wang2024starling, coleman2022graph, kang2025scalable, yue2025select}. Starling~\cite{wang2024starling} aims to place neighboring nodes in the same index block to reduce the total number of I/O calls performed during expansion; MARGO~\cite{yue2025select} builds on Starling to additionally place nodes on common paths into the same block aiming to perform multiple expansion rounds with one roundtrip to storage. According to our findings on bottlenecks of graph indexes when high beamwidth values are used (\cref{sec:exp_parameterization}), these works would effectively address incurred IOPS limits to further increase throughput under high concurrency and long roundtrip times, respectively, making their integration into on-cloud vector search pipelines promising.

\paragraph{Quantization for cluster indexes}
Some works have proposed quantization-aware optimizations for cluster indexes~\cite{gao2024rabitq, ivfpq, noh2021product, kalantidis2014locally}.
A common theme is to reduce the accuracy loss incurred on the index by quantization, which techniques such as clustering on quantized distances~\cite{noh2021product, gao2024rabitq} and per-posting list quantization~\cite{kalantidis2014locally} aim to accomplish. Orthogonally, IVF-PQ~\cite{ivfpq} partitions vectors into segments and performs per-segment quantization and clustering. We have observed that the two key bottlenecks of on-cloud cluster indexes are the large amount of data read and I/O calls (\cref{sec:exp_e2e}); while quantization would reduce posting list size and improve throughput in bandwidth-constrained settings, our results suggest that quantization would be less effective when the IOPS constraint is instead reached (e.g., on low-dimension datasets such as \deep, \cref{tbl:workload}), necessitating alternative techniques such as caching.

\paragraph{ANN query latency reduction via early stopping}
Studies have found vector ANN queries to feature varying hardness levels~\cite{angiulli2018behavior, wang2024steiner}, and works have accordingly proposed methods for early-stopping easy queries in both graph indexes~\cite{chen2021spann, li2020improving, zhang2023fast} and cluster indexes~\cite{xu2025tribase, song2025trim}. LAET~\cite{li2020improving} and Auncel~\cite{zhang2023fast} uses mid-query-serving features to dynamically early-stop graph traversal; SPANN~\cite{chen2021spann} uses a heuristic-based method to skip posting lists with centroids far from the query vector, while Tribase~\cite{xu2025tribase} and TRIM~\cite{song2025trim} further use triangle inequalities to early stop within-posting list distance computations. While early stopping will improve the performace of on-cloud search with graph indexes by reducing roundtrips, our results suggest that adopting within-posting list early stopping for cluster indexes will require more consideration as these techniques aim to reduce the computations on \textit{already fetched} data, which is often not a primary bottleneck of cluster indexes (\cref{sec:exp_parameterization}).

\paragraph{Caching for Vector Indexes}
GoVector~\cite{zhou2025govector} and CALL~\cite{jeong2025call} have recently proposed methods for improving the cache hit rate of caching strategies applied to graph index blocks and cluster index posting lists, respectively. GoVector utilizes a path-aware caching strategy to prioritize keeping nodes on the same common paths, while CALL proposes a grouping strategy to improve cache locality. Specifically for graph indexes, our work has shown that higher cache hit rate does not necessarily imply lower latency for individual queries when combined with \diskann's beamwidth-specified multi-node expansion (\cref{sec:exp_caching}), and that further developing a beamwidth-aware caching strategy would be valuable future work for on-cloud search.

% \begin{enumerate}
%     \item Open discussions\begin{enumerate}
%         \item 
%     \end{enumerate}
%     \item Optimizations for on-disk vector indexes\begin{enumerate}
%         \item Starling
%         \item SPFresh
%         \item IP-Diskann
%         \item Starling is relevant in exploring solutions to the problems with on-remote storage indexing we expose in this paper. The other works are mostly orthogonal, we don't discuss index updates in this paper
%     \end{enumerate}
%     \item On remote-storage vector indexing \begin{enumerate}
%         \item Turbopuffer
%         \item Turbopuffer uses a very vanilla setup, we provide insights for optimizations
%     \end{enumerate}
%     \item Caching for on remote-storage workloads \begin{enumerate}
%         \item Intermediate result reuse works
%         \item Cache eviction policy works
%         \item We provide insights on how caching may or may not be useful (and how it should be adapted) for vector search on remote indexing
%     \end{enumerate}
%     \item Remote storage optimizations\begin{enumerate}
%         \item TBD (maybe some works talking about optimizations w.r.t. hot/cold data, bandwidth, etc)
%     \end{enumerate}
% \end{enumerate}
\section{Conclusion}
In this paper, we present a comprehensive performance analysis for cloud-native vector search. We perform principled cost modeling for two representatives of the commonly-used graph and cluster vector index classes, \spann and \diskann, both theoretically and empirically show their bottlenecks in cloud-native vector search settings, and how local caching can be integrated with these indexes for performance gains. Then, we accordingly provide insights on (1) which index to use for a wide variety of workloads with different latency, throughput, recall, query concurrency, and dataset properties, (2) how to improve the performance of the chosen index via tuning its parameterization according to the workload characteristics, and (3) how indexing should be adjusted for further performance gains under various available local cache sizes.

% an indexing framework enabling efficient filtered vector search via an index collection. \system uses a three-dimensional cost model for memory size, search speed, and recall to determine benefits and costs of candidate indexes at a target recall to build the index collection with bounded memory via workload-driven optimization. For query serving, \system finds the fastest search strategy---a parameterized index search or brute-force KNN, at a potentially new target recall. \system achieves up to 8.06$\times$ QPS gain over existing indexing methods at 0.9 recall on various datasets while requiring as little as 1\% TTI versus other specialized indexes.

%%
%% The next two lines define the bibliography style to be used, and
%% the bibliography file.
\bibliographystyle{ACM-Reference-Format}
\bibliography{main}
% \appendix
% \subfile{sections/appendix}
\balance

\end{document}